\documentclass[screen, sigconf]{acmart}
\AtBeginDocument{%
  }

\copyrightyear{2026}
\acmYear{2026}
\setcopyright{cc}
\setcctype{by-nc-nd}
\acmConference[CHI '26]{Proceedings of the 2026 CHI Conference on Human Factors in Computing Systems}{April 13--17, 2026}{Barcelona, Spain}
\acmBooktitle{Proceedings of the 2026 CHI Conference on Human Factors in Computing Systems (CHI '26), April 13--17, 2026, Barcelona, Spain}
\acmPrice{}
\acmDOI{10.1145/3772318.3790792}
\acmISBN{979-8-4007-2278-3/2026/04}

\usepackage{pifont} 
\usepackage{fontawesome}
\usepackage{soul}
\usepackage{multirow}
\usepackage{float}
\usepackage{graphicx}
\usepackage[justification=centering]{subfig}
\usepackage{makecell}
\usepackage{booktabs}
\usepackage{stfloats}
\usepackage{enumitem}
\usepackage{subcaption}
\usepackage{xcolor}
\usepackage{color, colortbl}
\usepackage{rotating}
\usepackage{xspace}

\begin{document}

\newcommand\add{\added[id=add]}
\newcommand\del{\deleted[id=del]}
\newcommand\rep{\replaced[id=rep]}
\title{Red Teaming LLMs as Socio-Technical Practice: From Exploration and Data Creation to Evaluation}


\author{Adriana Alvarado Garcia}
\orcid{0000-0002-4230-3777}
\authornote{Both authors contributed equally to this research. This research was conducted during Ruyuan Wan’s internship at IBM Research.}
\affiliation{%
\department{Responsible Tech Research}
  \institution{IBM Research}
  \city{Yorktown Heights}
  \state{New York}
  \country{USA}}
\email{adriana.ag@ibm.com}

\author{Ruyuan Wan}
\authornotemark[1]
\orcid{0000-0002-0357-5139}
\affiliation{%
    \department{College of Information Sciences and Technology}
    \institution{The Pennsylvania State University}
    \city{University Park}
    \state{Pennsylvania}
  \country{USA}}
\email{rjw6289@psu.edu}

\author{Ozioma C. Oguine}
\orcid{0000-0003-2434-1400}
\affiliation{%
\department{Computer Science and Engineering}
  \institution{University of Notre Dame}
  \city{Notre Dame}
  \state{Indiana}
  \country{USA}}
\email{ooguine@nd.edu}

\author{Karla Badillo-Urquiola}
\orcid{0000-0002-1165-3619}
\affiliation{%
\department{Computer Science and Engineering}
  \institution{University of Notre Dame}
  \city{Notre Dame}
  \state{Indiana}
  \country{USA}}
\email{kbadill3@nd.edu}

\renewcommand{\shortauthors}{Alvarado Garcia and Wan, et al.}

\begin{abstract}
Recently, red teaming, with roots in security, has become a key evaluative approach to ensure the safety and reliability of Generative Artificial Intelligence. However, most existing work emphasizes technical benchmarks and attack success rates, leaving the socio-technical practices of how red teaming datasets are defined, created, and evaluated under-examined. Drawing on 22 interviews with practitioners who design and evaluate red teaming datasets, we examine the data practices and standards that underpin this work. Because adversarial datasets determine the scope and accuracy of model evaluations, they are critical artifacts for assessing potential harms from large language models. Our contributions are first, empirical evidence of practitioners conceptualizing red teaming and developing and evaluating red teaming datasets. Second, we reflect on how practitioners’ conceptualization of risk leads to overlooking the context, interaction type, and user specificity. We conclude with three opportunities for HCI researchers to expand the conceptualization and data practices for red-teaming.
\end{abstract}

\ccsdesc[500]{Human-centered computing~Empirical studies in HCI}
\ccsdesc[500]{Computing methodologies~Natural language generation}
\ccsdesc[300]{Security and privacy~Human and societal aspects of security and privacy}

\keywords{Red Teaming, Language Models, Evaluation, Data Practice, Data Work, Socio-Technical, Artificial Intelligence}


\maketitle

\section{Introduction}
Red teaming originates from the context of security, where adversaries simulate attacks to uncover vulnerabilities \cite{Sinha2025Jul, Feffer2024Oct, zenko2015red, Anderson2025Jul}. Nowadays, red teaming has become central to evaluating the safety and robustness of systems, especially with the rise of large language models (LLMs). LLMs now outperform benchmarks in reasoning and language understanding \cite{achiam2023gpt}, but also produce harmful, biased, or misleading outputs. These risks have made red teaming an emerging standard practice in industry, government, and academia, embedded in regulations such as the European Union’s Artificial Intelligence Act \cite{EUAIact}. 

Adversarial datasets are central to red teaming LLMs, revealing models’ vulnerabilities and blind spots. These datasets are collections of prompts or conversations designed to elicit harmful or unsafe behaviors, enabling systematic evaluation and alignment improvements.
They define what counts as harm, determine how models are tested, and shape what risks present to downstream users. Yet datasets are not neutral: they embody the assumptions and values of those who create them \cite{Miceli2020, Scheuerman2021,Sambasivan2021May}. Prior work in Human-Computer Interaction (HCI) and critical data studies has shown that annotation practices, labeling schemes, and benchmark structures not only shape technical outcomes but also social and cultural norms \cite{Miceli2020, Scheuerman2021,Sambasivan2021May}. Data workers develop intuitive understandings of their data through multiple forms of intervention, including treating data as given, captured, curated, designed, or created ~\cite{muller2019data}. However, little is known about how AI practitioners create, reuse, and evaluate red teaming datasets for LLMs. Different from many prior data work that analyze data produced in real-world contexts, such as social media content~\cite{alvarado2025knowledge} or image recognition~\cite{Miceli2020}, red-teaming datasets are created with an explicitly adversarial, experimental, and testing-oriented intention. Rather than capturing naturally occurring human activity, practitioners design prompts to provoke failures, explore attack strategies, and probe the boundaries of model behavior~\cite{lin2025against}. This adversarial stance reframes data work in this context: while red-teaming data creation is highly open-ended and unstructured, it still depends on human judgment about what to include and how to validate it. Most red teaming LLMs research emphasizes technical benchmarks and attack success rates, leaving the socio-technical practices of dataset development under-examined. These refer to the intertwined social and technical choices about what should be supported socially, and what can be implemented technically \cite{ackerman2000intellectual}, such as how harms are defined, how adversarial prompts or dialogue are created, and what evaluation metrics are adopted.

This gap matters because harmfulness is not discovered as a fixed concept in red teaming evaluation, but constructed through choices about prompts, categories, evaluation metrics, and collaborations. Decisions about what harms to include or exclude, and how to measure success, shape not only the coverage of datasets but also the guardrails and mitigations that follow. Understanding these practices is critical for HCI, which has long examined how technologies embed human values, and for AI safety, which increasingly depends on socio-technical insights.
To address this gap, we conducted 22 semi-structured interviews with AI practitioners who design, build, or reuse datasets for red teaming LLMs. Our study is guided by the following research questions:
\begin{enumerate}
    \item How do AI practitioners create, develop, and evaluate red teaming datasets, and why in these ways?
    \item What tools and support do AI practitioners need when developing red teaming datasets?
\end{enumerate}

Our work extends prior HCI examinations of data work by revealing a distinct form of adversarial, interaction-driven data work that emerges uniquely in red teaming LLMs.
Our analysis highlights three critical moments in this work: (1) defining and framing red teaming tasks, (2) developing adversarial datasets, and (3) evaluating models against them. 
Across these moments, we show that disciplinary backgrounds shape whether red teaming is framed as exploration or classification, while motivations range from observing ``in-the-wild'' jailbreaks to addressing technical or regulatory gaps. We find that datasets are developed through different pathways: created from scratch, repurposed from existing resources, or derived from human interactions, and that evaluation is complicated by issues of context, diversity, and metrics. Taken together, these practices reveal red teaming as more interactional and social than AI practitioners typically anticipated in benchmark-driven work, with risks arising not only from single prompts but also from multi-turn, multilingual, and multicultural exchanges. 
Further, we identify opportunities for HCI researchers to support red teaming practitioners: by expanding evaluations to center on the context of use, incorporating domain expertise into definitions of harm, and better accounting for interaction-level risks. 
Our findings shed light on emerging exploratory data practices that arise when LLMs are engaged through open-ended and interactive use.

\section{Related Work}
To contextualize our contribution, we first introduce the concept and practice of red teaming, then examine how recent datasets operationalize adversarial testing for LLMs. We then consider the diverse stakeholders involved in red teaming for AI safety, highlighting the motivations and tensions that shape its practice. Finally, we situate our work within HCI research, showing how longstanding contributions on data practices provide critical insights for understanding and improving red teaming evaluation.

\subsection{Introduction to Red Teaming} 
Red teaming did not originate as a technical protocol, but rather as a strategic exercise in adversarial thinking. Rooted in military war-gaming, red teams were tasked with challenging institutional assumptions by “\textit{thinking like the enemy}” to expose vulnerabilities in operational plans \cite{zenko2015red, Anderson2025Jul, Sinha2025Jul}. Over time, this practice migrated into cybersecurity, where red teams simulate breaches to identify flaws in digital defenses, a tradition now formalized in frameworks like the U.S. Department of Defense’s Cyber Red Teaming Guide \cite{Sinha2025Jul, Feffer2024Oct}. This adversarial framing of treating safety as a matter of defense against active threats shaped how risk was conceptualized in digital systems, establishing a dominant view of harm as something imposed by hostile actors, rather than as the emergent, everyday vulnerabilities experienced by diverse users \cite{Feffer2024Oct, Ahmad2025openai}. It provided a blueprint for how harms are anticipated, operationalized, and mitigated in sociotechnical design \cite{Gillespie2024Dec, Solaiman2023Jun}. Academic scholarship across security studies \cite{mansfield2018best}, information systems \cite{Applebaum2016redteam}, CSCW and HCI ~\cite{Zhang2025Jun, zhang2024human, zhang2025work} contributed to this transition, theorizing red teaming as a method of system auditing, threat modeling, and scenario-based testing. These foundational contributions laid the groundwork for the widespread adoption of red teaming in industry, government, and research settings.

In the domain of AI, and particularly with the rise of LLMs and emerging agentic AI systems, red teaming has taken on renewed urgency \cite{Weidinger2023eval, Ahmad2025openai, Feffer2024Oct}. While LLMs offer substantial societal benefits, including expanded access to information, enhanced productivity, and support for human creative and cognitive tasks~\cite{He2024AI-collaborate, suh2024luminate, zhang2025ladica}, they have also been shown to generate outputs that are toxic, biased, or sometimes inaccurate \cite{Weidinger2021harms}. In response, the Center for AI Safety released a statement in 2023, warning that AI extinction risk should be a global priority \cite{cais2023statement}. Therefore, governments (e.g., the EU's AI Act \cite{EUAIact} and the U.S. Executive Order 14110 \cite{Biden2023AI}) and industry leaders (e.g., \citet{openAI2024}, \citet{anthropic2023ModelCard}) have institutionalized red teaming as a key part of AI evaluation.
While academic researchers have also emphasized the need for rigorous evaluation frameworks to ensure the safety and reliability of AI systems in real-world settings (e.g., \cite{Majumdar2025Jul, Weidinger2023eval, Perez2022Feb, Perez2022Feb, Ganguli2022Aug, solyst2023bias, Feffer2024Oct, Morales2024auditor}), their valuable contributions are often overshadowed by industry-led initiatives. 

Red teaming has emerged as a field of study connecting security, AI safety, and responsible innovation. Yet, the conceptual framing of red teaming remains largely inherited from its adversarial roots \cite{Feffer2024Oct}. The dominant notion of a “threat” is typically cast as a malicious actor or unexpected query, someone trying to game or break the system \cite{Feffer2024Oct, zenko2015red}. This framing often sidelines more systemic or structural harms, such as how LLMs may reinforce racial stereotypes, marginalize dialects, or exclude non-Western epistemologies \cite{Gillespie2024Dec, singh2025red}. As a result, current red teaming practices tend to reflect narrow slices of AI risk: those that are easily measurable, align with regulatory compliance, or reflect high-profile public controversies. 
By centering the people and practices involved, we showed that red teaming--often framed as technical safety work--is shaped by social values and disciplinary priorities.

\subsection{The Role of Red Teaming Datasets for LLMs}
A growing body of research has demonstrated that LLMs can produce harmful, toxic, or biased outputs, even after alignment efforts \cite{Gehman2020Toxicity, Weidinger2021harms, Perez2022Feb, Ganguli2022Aug}. For example, both \citet{Gehman2020Toxicity} and \citet{Welbl2021Nov}, documented how LLMs can produce toxic and misleading outputs even under controlled input conditions. Early generative models like GPT-2 responded with toxic completions to 4.3–6.1\% of user prompts, depending on prompt structure and sampling temperature \cite{Gehman2020Toxicity}. \citet{Sheng2019Nov} also showed that models can reinforce gender and racial stereotypes in generated sentences. These findings sparked renewed attention to the limitations of existing evaluation pipelines and the need for targeted adversarial testing.

In response, researchers have turned to red teaming datasets, structured collections of adversarial or high-risk prompts used to probe model behavior \cite{Ganguli2022Aug, Perez2022Feb, Feffer2024Oct}. Unlike benchmark datasets designed to test factual accuracy or task performance \cite{Wang2018Nov, Rajpurkar2016Nov}, red teaming datasets are crafted to elicit failure modes, especially in high-risk domains like hate speech, and safety-critical advice \cite{Ganguli2022Aug, Perez2022Feb}. These datasets act as epistemic instruments, shaping which types of harms are visible, how safety is defined, and what mitigation efforts follow \cite{Feffer2024Oct, Sambasivan2021May, AlvaradoGarcia2025Apr, Muller2022forgetting}. An early impactful effort, RealToxicityPrompts by \citet{Gehman2020Toxicity}, included 100,000 prompts mined from web text that elicited toxic completions from pre-trained models. Each prompt was annotated using the PerspectiveAPI to measure the severity of model toxicity, forming a baseline for assessing generative harm. Yet as we discuss later, reliance on automated toxicity scores also obscures questions of diversity and cultural context in how harm is recognized. 
\citet{Liang2022Nov} introduced HELM, which evaluates LLMs across 42 metrics spanning bias, toxicity, factuality, and robustness. Although HELM is not exclusively adversarial, it demonstrated how evaluation pipelines increasingly include risk-centric benchmarks. Recent red teaming datasets such as HH-RLHF \cite{Ganguli2022Aug}, AdvBench \cite{Zou2023universal}, HarmBench \cite{Mazeika2024harmbench}, and Beavertails \cite{Ji2023Beavertails} further illustrate this shift, embedding adversarial evaluation directly into model development pipelines and broadening the range of harms under examination beyond traditional performance metrics. \autoref{tab:redteaming-datasets} compares these four widely used datasets to highlight how differences in their size, generation methods, harm definitions, and design inspirations shape what risks are surfaced and which may remain overlooked in red teaming practice.

\begin{table*}[htbp]
\scriptsize
\newcolumntype{L}[1]{>{\raggedright\arraybackslash}p{#1}}
\begin{tabular}{|L{1.6cm}|L{0.5cm}|L{1.5cm}|L{2.4cm}|L{1.1cm}|L{1.2cm}|L{3.3cm}|}
\hline
\textbf{Dataset} & \textbf{Year} & \textbf{Size / Type} & \textbf{Design Basis} & \textbf{Generation Method} & \textbf{Annotation Method} & \textbf{Definition of Harm} \\
\hline
HH-RLHF \cite{Ganguli2022Aug} & 2022 & $\sim$38k human-generated comparisons & Anthropic's “helpful/harmless” framework & 
\raisebox{-0.6\height}{\includegraphics[height=1.5em]{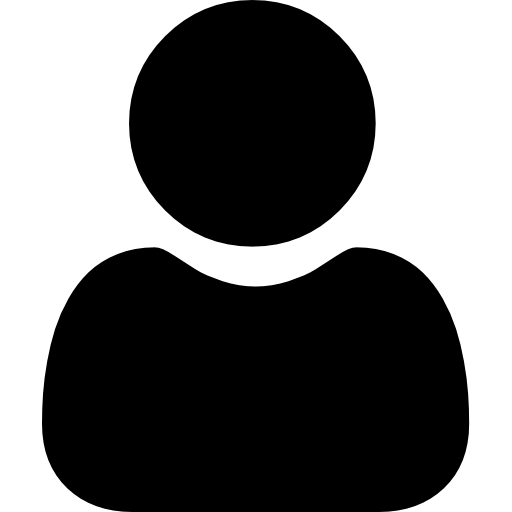}} & 
\raisebox{-0.6\height}{\includegraphics[height=1.5em]{Images/user.png}} & 
Harm defined by human-preference for ``hamrful'' outputs \\
\hline
AdvBench \cite{Zou2023universal} & 2023 & 500 adversarial “strings” & Adversarial robustness testing focused on policy-violating behavior& 
\raisebox{-0.6\height}{\includegraphics[height=1.5em]{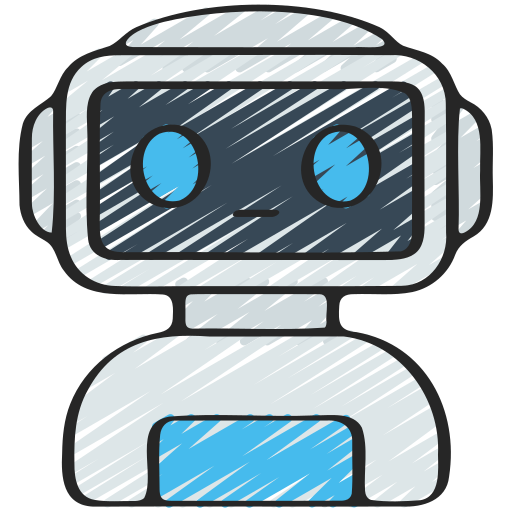}} & 
\raisebox{-0.6\height}{\includegraphics[height=1.5em]{Images/chatgpt.png}} & 
Harm framed as violating model’s alignment or safety constraints \\
\hline
BeaverTails \cite{Ji2023Beavertails} & 2023 & 30,207 Question–Answer pairs & OpenAI jailbreak prompt sets + community red teaming & 
\raisebox{-0.6\height}{\includegraphics[height=1.5em]{Images/chatgpt.png}} & 
\raisebox{-0.6\height}{\includegraphics[height=1.5em]{Images/chatgpt.png}} \raisebox{-0.6\height}{\includegraphics[height=0.7em]{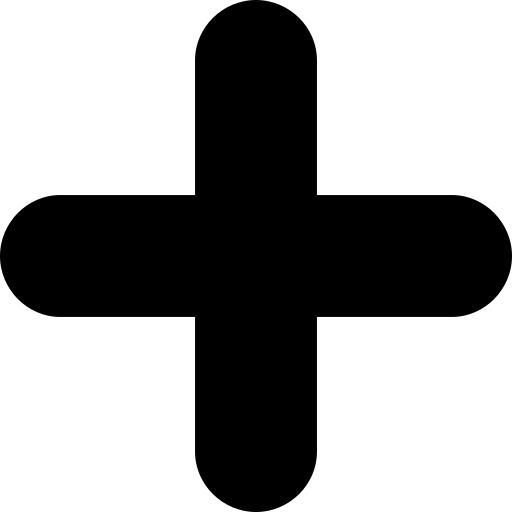}} \raisebox{-0.6\height}{\includegraphics[height=1.5em]{Images/user.png}} & 
Harm defined in relation to jailbreak success and unsafe generations \\
\hline
HarmBench \cite{Mazeika2024harmbench} & 2024 & $\sim$2,400 test cases & Prior harm taxonomies + academic harm frameworks & 
\raisebox{-0.6\height}{\includegraphics[height=1.5em]{Images/user.png}} \raisebox{-0.6\height}{\includegraphics[height=0.7em]{Images/add.png}} \raisebox{-0.6\height}{\includegraphics[height=1.5em]{Images/chatgpt.png}} & 
\raisebox{-0.6\height}{\includegraphics[height=1.5em]{Images/user.png}} \raisebox{-0.6\height}{\includegraphics[height=0.7em]{Images/add.png}} \raisebox{-0.6\height}{\includegraphics[height=1.5em]{Images/chatgpt.png}} & 
Harm defined by standardized taxonomy (violence, hate, self-harm, etc.) \\
\hline
\end{tabular}

\vspace{0.5em}
\scriptsize{\textbf{Legend:} \raisebox{-0.6\height}{\includegraphics[height=1.5em]{Images/user.png}} = Human; \raisebox{-0.6\height}{\includegraphics[height=1.5em]{Images/chatgpt.png}} = AI/LLM; 
\raisebox{-0.6\height}{\includegraphics[height=1.5em]{Images/user.png}} \raisebox{-0.6\height}{\includegraphics[height=0.7em]{Images/add.png}} \raisebox{-0.6\height}{\includegraphics[height=1.5em]{Images/chatgpt.png}} = Human-AI collaboration}
\caption{Comparison of major Red Teaming Datasets for LLMs}
\label{tab:redteaming-datasets} 
\end{table*}

Despite their importance, red teaming datasets are far from neutral. As \citet{Solaiman2023Jun} argued, decisions around prompt selection, annotation frameworks, and harm taxonomies reflect underlying social and institutional values. These are not purely technical design choices, but data practices, cultural, political, and labor-intensive processes that shape how risk is constructed and operationalized \cite{Gillespie2024Dec,Scheuerman2021,Miceli2020}. Consistent with prior work, AI datasets including those used in red teaming, are overwhelmingly created by institutions in the Global North, with minimal inclusion of perspectives from the Global South or marginalized communities \cite{Sambasivan2021May,Costanza-Chock2020Mar}. Furthermore, most datasets rely on English-language prompts, limiting their ability to capture harms experienced by multilingual or culturally diverse communities \cite{ramesh-etal-2023-fairness}.

These data practices have tangible consequences for the visibility and mitigation of harm. If red teaming datasets fail to include prompts reflecting LGBTQ+ experiences, youth-specific concerns, or migrant community perspectives, then these harms may remain unexamined \cite{Weidinger2021harms}. \citet{Hofmann2024Sep} for example, found that LLMs assigned more negative stereotypes, lower‑prestige jobs, and harsher criminal judgments to speakers of African‑American English compared to Standard American English prompts highlighting how dialect bias can shape harmful system behavior. By focusing on red teaming as a site of data work, our study investigates how practitioners make decisions about which risks to include, what constitutes a “good” prompt, and how these choices influence downstream definitions of safety and model accountability.

\subsection{Stakeholders in Red Teaming for AI Safety} 
While red teaming is often framed as a technical practice aimed at detecting model vulnerabilities, it is in fact deeply shaped by the social, institutional, and political context in which it is conducted \cite{Gillespie2024Dec,singh2025red}. A growing body of research has highlighted that the definition of “risk” in red teaming is not universal, but reflects the values, priorities, and threat models of those with the power to define what is tested and how \cite{Feffer2024Oct, Gillespie2024Dec, Weidinger2023eval,singh2025red}. As such, understanding red teaming requires attending to its stakeholders: the diverse and sometimes conflicting set of actors involved in shaping, executing, and responding to safety evaluations. These stakeholders include AI developers, research scientists, policy regulators, product managers, red teamers, annotators, and users; each bringing different motivations, expertise, and constraints \cite{schwartz2025, singh2025red}. Developers often prioritize model alignment and technical reproducibility, while policy teams focus on regulatory compliance or reputational risks \cite{Weidinger2021harms, Feffer2024Oct}. End users -- particularly those from marginalized communities -- may be more concerned with harms grounded in lived experience, such as misgendering, cultural erasure, or coded hate speech, which are rarely formalized in red teaming pipelines 
~\cite{Costanza-Chock2020Mar, wan2023community}. 
These divergent priorities create tensions over whose risks are surfaced and whose are overlooked \cite{Gillespie2024Dec, Selbst2019Jan}.

Despite growing calls for inclusive safety evaluation by several scholars \cite{delgado2023participate, Longpre2024Mar}, the perspectives of affected communities are often absent or tokenized in practice. Also, current red teaming frameworks lack clear mechanisms for how and where to integrate these perspectives \cite{Kallina2025Jun, oguine2025bridging}. As \citet{delgado2023participate} highlighted, participatory evaluation processes tend to be extractive, consulting users or advocacy groups without redistributing power over decision-making or interpretation. This raises critical questions about what forms of expertise are legitimized in red teaming and what it means to evaluate a model “responsibly.” Recent work has begun to surface these participatory gaps \cite{solyst2023bias, delgado2023participate, Morales2024auditor}. \citet{Gillespie2024Dec} frames red teaming as a sociotechnical system in which the labor, institutional incentives, and cultural assumptions of developers profoundly shape which harms are tested. \citet{Zhang2025Jun} shows how data workers involved in curating red teaming prompts must make judgment calls without clear guidance or support, revealing the limits of centralized governance. Together, these studies suggest that stakeholder inclusion is not just a matter of ethics but a necessary condition for meaningful safety. When the definition of harm is negotiated in closed circles, it risks being narrow, reactive, and disconnected from the realities of deployment.  

Our study builds on this body of work by examining how red teaming practitioners reflect on stakeholder roles: who is consulted, who is represented, and how inclusion is operationalized or not during the design of red teaming datasets. 
From these perspectives, we surface the hidden practices and trade-offs that decide which harms are recognized and which are ignored in safety evaluation.

\subsection{Contribution of HCI to Red Teaming Evaluation} 
Scholars within the SIGCHI community have long studied the sociotechnical nature of system evaluation and more recently red teaming \cite{Zhang2025Jun,zhang2024human, Xiao2024audit, lam2023sociotechnical}, offering critical insights and tools for understanding how values, labor, and context shape technical processes. In contrast to dominant narratives in AI red teaming, which often emphasize technical rigor, adversarial testing, and benchmark-based safety, HCI scholarship foregrounds the social dimensions of harm \cite{saxena2025ai}, the labor of dataset creation \cite{AlvaradoGarcia2025Apr, Sambasivan2021May}, and the need for participatory, situated evaluation \cite{ma2025sphere}. 
This work highlights red teaming not only as a way of testing models, but as a practice that defines harm and safety.

Additionally, foundational insight from HCI research emphasizes that data work, such as curating, cleaning, labeling, and evaluating datasets, is not neutral or mechanistic, but socially embedded and judgment-laden. Studies have shown that practitioners making decisions about datasets often operate with little recognition, institutional support, or shared criteria \cite {Qian2024May, Sambasivan2021May}. These decisions, however, have far-reaching effects: they shape how systems behave, how failures are detected, and how “responsible AI” is defined in practice \cite{AlvaradoGarcia2025Apr, Gillespie2024Dec}. Red teaming, viewed through this lens, is a form of high-stakes data work that involves implicit negotiations around value trade-offs, stakeholder representation, and institutional norms.

HCI studies have also contributed methodological alternatives to \textit{opaque-box} auditing or static red teaming pipelines. Scholars have explored participatory approaches to algorithmic evaluation that involve users and affected communities in defining harms and co-constructing risk scenarios \cite{Costanza-Chock2020Mar}. These approaches push against the abstraction and decontextualization common in much of AI evaluation. For instance, participatory design methods in HCI have long emphasized that “designing for safety” requires accounting for context, lived experience, and systemic power, not just adversarial queries or edge cases \cite{Badillo-Urquiola2024Apr, Oguine2025Apr}. Moreover, HCI critiques the assumption that evaluation can ever be value-neutral. As \citet{Selbst2019Jan} argued, technical interventions often embed unstated normative assumptions about what kinds of errors are tolerable, whose experiences count as harm, and what trade-offs are acceptable. This critique is directly relevant to red teaming, where prompts are often framed as objective “tests,” though their design reflects deeply contextual choices \cite{Wong2017Dec, Sambasivan2021May}. These choices are rarely documented, contested, or made transparent \cite{Feffer2024Oct, Muller2022forgetting}.

Drawing from this HCI lineage, our study treats red teaming not as a purely technical task but as a sociotechnical practice involving meaning-making, boundary-drawing, and negotiation, in partnership with the people most affected by these decisions. Further, we call for more participatory, transparent, and reflexive approaches to evaluating AI systems that center not only on what a model can do, but who decides what it should do, and for whom.

\section{Methods}
In this section, we provide an overview of our interview study. We describe the recruitment strategy we follow to ensure that we interview practitioners with red teaming expertise, and provide a characterization of the participants' profiles. We conclude with a description of our data analysis process. 

\subsection{Recruitment} 
To understand how red teaming datasets are designed, developed, and evaluated, we recruit AI practitioners who have released public red teaming datasets or have experience researching red teaming datasets. 
Our process consisted of first identifying potential participants with experience in red teaming datasets for LLMs. We searched for red teaming datasets from Hugging Face \footnote{https://huggingface.co/} and Papers with Code platforms \footnote{https://paperswithcode.com/}. We searched `red teaming' in full-text search in Hugging Face to collect red teaming datasets and the data creators’ information. We stopped after the top 100 search results. Because in the latter portion of the 100, there are many duplicated datasets from higher-ranked original red teaming datasets. Then we searched `red teaming' on Papers with Code, which gave us 100 results of papers and their GitHub repositories. We also explored additional communities, such as Kaggle, OpenML, and the Data Provenance Initiative, but only found five red teaming datasets\footnote{Kaggle had four, the Data Provenance Initiative had one, and OpenML had none.} which we already collected through our search in the Hugging Face and Paper with Code platforms. 

After gathering the red teaming datasets, we removed duplicates and selected potential participants following the criteria: the work should focus on natural language tasks, excluding efforts centered on multi-modal tasks, mathematical problem solving, security exploits, or code generation. The potential participants should have worked on red teaming projects for the purpose of developing, evaluating, or improving LLMs, interactive agents, and applied language technologies. This includes those who had created or curated red teaming prompts and datasets, designed pipelines for generating preference-aligned or adversarial data. After screening the search results, we recorded 107 qualified potential participants from open source communities, startups, industry, and academia.  

For each potential participant, we recorded the following information to guide our recruitment process: the name of the dataset, its associated link, the title and link of the corresponding paper, the primary practitioner's name, email, and affiliation. We also noted public engagement metrics of the red teaming datasets, including the number of last month's dataset downloads on Hugging Face, GitHub stars, and paper citation counts. Based on this information, we synthesized two categories of data practices, which informed our sampling strategy and interview targeting: 
\begin{enumerate}
    \item Create red teaming dataset from scratch: design and create original data specifically for red teaming. Practitioners can maximize the control and ownership of the dataset, but require more resource investment.
    \item Use existing dataset without extra revision: practitioners directly apply existing datasets to their studies. In doing so, they worked within the structure, assumptions, and classification choices embedded in the data, which reflect the values of the original dataset creators \cite{Miceli2020}, and the downstream cascading effects of data quality in high-stakes AI applications~\cite{Sambasivan2021May}.
\end{enumerate}

We ranked the potential participants based on three indicators of visibility and impact: the number of dataset downloads on Hugging Face or Papers With Code, the number of GitHub stars for associated repositories, and the number of citations for corresponding papers. From this ranked list, we sampled participants evenly across the two data practice categories and invited 76 potential participants in batches, starting from the highest-ranked. In total, 22 participants accepted our invitation: twelve created their own datasets and ten reused existing datasets. We stopped the recruitment process once we reached data saturation.

\subsection{Participant's Profiles}
Our participants are AI practitioners in eight countries on three continents. Fifteen are PhD students, and two have worked on red teaming projects during their research internships. All participants had a computer science background. We recorded their areas of expertise based on their self-description. While some descriptions were similar, they contained nuanced differences. Fourteen described having expertise on language models, LLMs, or NLP; seven focused on safety or security; and four participants identified their background as machine learning or reinforcement learning. 
Most of the practitioners contributing to red teaming LLMs came from academia, reflecting the trend we observed in our recruitment. From the 107 datasets and repositories we identified, 39 originated from academia alone, 34 from academia-industry collaborations, while only 19 were solely from industry and 15 from open-source communities. Among the 76 invitations we eventually sent, there were 16 industry practitioners, of which only three agreed to participate in our interview (14\% of 22 interviews).  
Details of our participants' profiles are shown in the Table \ref{tab:participants}.

\begin{table*}[htbp]
\centering
\begin{tabular}{ c l l l l l}
\toprule
\textbf{ID} & \textbf{Location} & \textbf{Background} & \textbf{Area of Expertise} & \textbf{Data Practice} & \textbf{Duration} \\ \midrule
P1  & Israel &  Research Scientist & Machine Learning &  Create  & 48 minutes\\ 
P2  & U.S. & Ph.D. Student & LLMs and Safety &  Reuse & 49 minutes \\ 
P3  & U.S. & Ph.D. Student & AI safety and security & Create & 53 minutes \\ 
P4  & U.S. & Ph.D. Student & Safety Alignment of LLMs & Reuse & 40 minutes \\ 
P5  & Singapore & Ph.D. Student & NLP &  Reuse & 44 minutes\\ 
P6  & U.S. & Ph.D. Student & Reinforcement learning &  Create & 51 minutes \\ 
P7  & Germany & Ph.D. Student & AI safety and LLMs &  Reuse & 56 minutes \\ 
P8  & Singapore & Postdoc & Machine learning, LLM safety & Reuse & 59 minutes\\ 
P9  & U.S. & Ph.D. Student & LLMs alignment & Reuse & 57 minutes\\
P10 & India & Undergrad & Security and AI security &  Create & 56 minutes\\ 
P11 & U.S. & Ph.D. Student & Reinforcement learning &  Reuse & 44 minutes\\ 
P12 & India & Entrepreneur & Software engineer &  Create & 47 minutes \\ 
P13 & Spain & Ph.D. Student & Testing language model & Create  & 51 minutes\\ 
P14 & China & High School Student & Computer Science &  Create & 59 minutes \\ 
P15 & Singapore &  Research Engineer &Software testing &  Create & 41 minutes\\ 
P16 & U.S. & Ph.D. Student & Language model and NLP &  Create & 45 minutes\\ 
P17 & South Korea & Ph.D. Student & Data augumentation and NLP & Create & 40 minutes\\ 
P18 & China & Ph.D. Student & Security of large models & Reuse & 57 minutes\\ 
P19 & U.S. & Ph.D. Student & NLP &  Create & 40 minutes \\ 
P20 & U.S. & Ph.D. Student & LLM reasoning &  Create & 52 minutes\\ 
P21 & South Korea &  Research Engineer & NLP, hate speech &  Reuse & 60 minutes\\ 
P22 & South Korea & Ph.D. Student & Efficiency in LLM &  Reuse & 60 minutes\\ 
\bottomrule
\end{tabular}
\caption{Participants' profile information and interview duration}
\label{tab:participants}
\end{table*}

\subsection{Interview Study}
We conducted 22 semi-structured interviews with AI practitioners, and each interview took between 40 to 60 minutes. Nineteen interviews were conducted in English, two were in Chinese, and one was in Spanish. The study was approved by the authors’ Institutional Review Board. Participation was voluntary. We informed the participants that all information they shared would be kept confidential and anonymous, and they permitted us to record the interview for further analysis. 

We chose semi-structured interviews as our primary method because they offer an appropriate way to understand practitioners' rationales when developing red teaming techniques and provide critical access to settings where ethnographic entry is constrained. 
Opportunities for fieldwork or ethnographic studies are limited across both industry and academia: red-teaming activities in industry often occur within proprietary or confidential environments \cite{Bajema_2024}, while academic red-teaming projects are also typically confidential until publication. 
Therefore, we consider interviewing as an appropriate mode of inquiry to understand practitioners' day-to-day activities that otherwise would remain inaccessible. 
Interviews also enabled us to reach participants working across diverse geographic and institutional contexts, which would have been logistically challenging to access through site-based fieldwork.

While we recognize that observational methods could help uncover the tacit and embodied aspects of data work, they limit the ability to surface the rationales, trade-offs, constraints, and assumptions that guide people's choices. In contrast, semi-structured interviews allow for eliciting accounts of not only what people do but why they do it. This is particularly important for examining the assumptions embedded in dataset construction and evaluation practices, which are often invisible in observational inquiry alone.

To reduce idealized recall, we grounded our interview protocols in participants’ publicly released datasets and articles focused on red teaming techniques. We asked about the challenges and limitations they experienced, and asked them to explain their practice decisions in their work. Before each interview, we reviewed the participants' published papers and Hugging Face dataset cards, if applicable, to familiarize ourselves with their work and prepare our interview questions. We started the interviews by learning the participants' experiences in red teaming. Then we went deeper to discuss participants' experiences in creating or reusing red teaming datasets, their motivation for red teaming research, their categorization of harm and risks, and evaluation approaches. For each interview, we tailored, added, or skipped some questions towards the participant's own experiences and insights.

\subsection{Transcript Analysis}  
The two authors who conducted the interviews also analyzed the transcripts using a thematic analysis approach \cite{braun2006using}. Both authors attended all 22 interviews, with each author leading and coding half of the sessions.
After coding the first few interviews independently, each author developed a preliminary codebook. We then met to compare two codebooks, discuss discrepancies, and merge them into a single version with agreement. Using this shared codebook, we coded the remaining transcripts, revisiting earlier transcripts when necessary to ensure consistency. We discussed ambiguous cases during the coding process and refined definitions as needed, resolving disagreements through discussion until consensus was reached. The final analysis produced 47 sub-codes grouped under 21 parent codes, capturing participants’ definitions of red teaming, their motivations, practices, and perspectives on the evolution of LLMs. The Table ~\ref{tab:themes_codes} shows the summary of the derived themes, codes, and sub-codes. 

\renewcommand{\arraystretch}{1.15}
\setlength{\tabcolsep}{5pt}
\begin{table*}[htbp]
\centering
\begin{tabular}{p{2.5cm} p{4cm} p{6cm}}
\toprule
\textbf{Theme} & \textbf{Code} & \textbf{Sub-codes} \\
\midrule

\multirow{7}{=}{Participants’ conceptualization of red teaming LLMs}
& Definitions of red teaming & How participants define and conceptualize red teaming; Blind spots of red teaming \\ \cmidrule(l){2-3}
& Participant’s background and experience & Education; Years of experience; Affiliation \\ \cmidrule(l){2-3}
& Participant’s context and motivations & Work context; Motivations for red teaming \\\cmidrule(l){2-3}
& Participant’s approach to red teaming & Methods and strategies; Experiences described \\\cmidrule(l){2-3}
& Challenges & Practical challenges \\\cmidrule(l){2-3}
& References & References to additional materials  \\\cmidrule(l){2-3}
& Impact & Beneficiaries; Future impact \\
\midrule

\multirow{5}{=}{Developing adversarial datasets}
& Tailoring datasets & Adaptation processes; Risk categories; Representation; Contextual relevance; Diversity \\\cmidrule(l){2-3}
& High quality data criteria & Definitions and standards for data quality \\\cmidrule(l){2-3}
& Seed data & Sources; Selection criteria \\\cmidrule(l){2-3}
& Harmfulness definition & Definitions and examples of harmfulness \\\cmidrule(l){2-3}
& Reuse and repurpose existing datasets & Decision-making process; Advantages; Limitations \\
\midrule

\multirow{9}{=}{Evaluating adversarial datasets}
& Machine evaluation & LLM judges; Classifiers; Limitations \\\cmidrule(l){2-3}
& Human evaluation & Annotation; Validation; Challenges \\\cmidrule(l){2-3}
& Evaluation (overall) & Dimension ranking; Cross-category performance; Infrastructure; Challenges\\\cmidrule(l){2-3}
& Standards & Use of guidelines; Customized evaluation; Limitations \\\cmidrule(l){2-3}
& Human vs. machine & Comparisons between human and machine evaluations \\\cmidrule(l){2-3}
& Metrics & Metrics used; Trade-offs \\\cmidrule(l){2-3}
& Experts and stakeholders & Collaborations; ; Expert evaluation  \\\cmidrule(l){2-3}
& Geographies and representation & Geo-cultural bias; Representation issues \\\cmidrule(l){2-3}
& Misclassification & Examples; Mitigation strategies \\
\bottomrule
\end{tabular}
\caption{Themes, codes, and sub-codes from the thematic analysis of interviews.}
\label{tab:themes_codes}
\end{table*}

\subsection{Positionality}
All authors have received formal training in HCI. The two lead authors bring complementary expertise: one in critical data studies, which examines how data is produced through specific social processes and practices \cite{iliadis2016critical}, and the other in the intersection of HCI and NLP, with expertise in subjective data annotation, interpretation, and evaluation of language model outputs. These perspectives shaped our study design, the framing of our questions, and our interpretation of participants’ experience and insights to both the technical and sociocultural dimensions of red teaming \cite{holmes2020researcher}. Two co-authors bring domain expertise in online safety, particularly around youth risk and sociotechnical harms. These perspectives shaped our critical attention to how red teaming datasets define harm, surface risk, and reflect the experiences of vulnerable users.

\section{Findings}
While red teaming is often framed as a technical practice aimed at detecting model vulnerabilities, studies have shown that the definition of ``risk'' is not universal but reflects the values, priorities, and threat models of those in charge of defining what is tested and how \cite{singh2025red, Feffer2024Oct, Gillespie2024Dec}. We observed this dynamic in practice. Our findings are organized into three sections, illustrating how practitioners define and operationalize risks across three moments in red teaming LLMs: (1) conceptualizing what red teaming means, (2) developing adversarial datasets to capture those risks, and (3) evaluating the outcomes of red teaming LLMs.

\subsection{Participants' conceptualization of Red Teaming LLMs}
\label{findings_1}
In this section, we illustrated how practitioners conceptualize red teaming LLMs by examining their definitions of red teaming, their motivations for red teaming LLMs, and their views of red teaming as an interactional and social process.

\subsubsection{Participants' definitions of red teaming LLMs} 
Most participants started working on red teaming LLMs around the time LLMs became widely accessible and began outperforming standard benchmarks in commonsense reasoning and language understanding \cite{achiam2023gpt}, raising concerns about safety and misuse. 
Six participants had worked on red teaming for about two years, seven had one year of experience, and many had just started for a few months. These marked the release of ChatGPT in 2022 as a turning point that raised people’s concerns about AI safety. As P8 shared, \textit{``Red teaming (LLMs) hasn’t been around for long ... Its appearance came with ChatGPT, which launched at the end of 2022. Although the field is very active now, its lifespan has only been about a year and a half.''} 
Learning from participants, we noticed that the work on red teaming LLMs emerged quickly in response to the fast-growing capabilities and safety concerns of LLMs. 

Researchers from various fields, including security, reinforcement learning, and NLP, worked on it with different assumptions and methodologies. From our study, we identified two major ways of conceptualizing red teaming LLMs, which were shaped by practitioners' disciplinary training and research backgrounds. 

(1) Exploration and searching from the reinforcement learning framing (N=20). Participants with reinforcement learning training approached red teaming as an exploration problem, where the task was to navigate a vast prompt space to uncover successful attack cases. P6 explained how red teaming is a hard exploration challenge.
\begin{quote}
    \textit{
    ``Red teaming is a very hard exploration problem because you can think of the space, the search space of the prompt is like all combinations of the possible text, right? And this is a giant space that you will never be able to. Uh, traverse. So, how to design a strategy to explore this space? Yeah, it's a hard problem.''
    } - P06
\end{quote}
In this framing, assumptions shifted away from predefined ground truths: what mattered was not labels but the efficiency of search. Accordingly, the methodological practices emphasized Attack Success Rate (ASR), coverage, and diversity as primary evaluation criteria. As P11 described,
\textit{``I think coverage like you're trying to cover it, cover the whole space of possible queries. That's one of the main, I think one of the most important in red teaming.''} This view aligns with the broad research community of red teaming generative models. \citet{lin2025against}'s survey also highlights that crafting diverse red teaming prompts requires creativity and labor effort, so many studies explore automatic methods to create such prompts with a defined state space, search goal, and search operations. Participants (P6, P11, P17, P19) noted that coverage without diversity risks redundancy, since many prompts may differ by just a few words. To address this, they emphasized balancing coverage with diversity, while others, such as P22, highlighted that efficiency also matters in query searching, especially given the API costs of large-scale experiments.

(2) Inappropriate content detection from the NLP framing (N=2). Participants with NLP backgrounds conceptualized red teaming through the lens of familiar classification tasks. In this view, red teaming was about detecting contradictions in dialogue, or identifying offensive content, which are traditional NLP problems. P16, for example, described red teaming as triggering self-contradictory responses across dialogue turns, reflecting the structure of dialogue detection tasks.
\begin{quote}
    \textit{``So I use this concept that red teaming is about using one language model to attack another model to trigger something. That's another model to answer something that makes it contradictory (content), you can consider it like two models, which we can call two agents. One is to ask another one to generate some contradictory, model safety, that is kind of the red teaming.''
    } - P16 
\end{quote}
P21 similarly framed red teaming as users intentionally tricking a chatbot into producing hate speech, aligning the work with established hate speech detection approaches.
\begin{quote}
    \textit{
    ``Our work is in between those two categories (hate speech detection and exploring potential jailbreaking prompt problem). This paper doesn't deal with those technological red teaming. But we framed this work as red teaming because we could find the users attempt to jailbreak this system (chatbot), which can also be one type of jailbreak that is manual... So from that perspective, this can be a red teaming work, but from the traditional hate speech perspective.''
    } - P21 
\end{quote}
Because of this framing, evaluation was tied to predefined ground truth labels of harm and measured through supervised metrics such as the F1 score.

These two conceptualizations reveal a fundamental contrast in how red teaming LLMs was understood. The exploration and search framing treated red teaming as an open-ended problem of discovering failures across an unbounded prompt space, where coverage, diversity, and efficiency were key. In contrast, the inappropriate content detection framing treated red teaming as a supervised classification problem, where harmfulness was predefined and measured against labeled ground truth. These practices illustrate how disciplinary backgrounds shaped practitioners’ assumptions about red teaming and guided their approaches.


\subsubsection{Motivation of Red Teaming LLMs}
Participants were motivated to study red teaming LLMs for diverse reasons, including (1) observing red teaming ``in the wild,'' (2) filling technical gaps, (3) responding to organizational influences, and (4) improving models' safety after red teaming.

A common motivation stemmed from observing users circumvent LLM safeguards and share their conversations online. For example, P12 became interested in the topic after seeing Reddit users share their conversations with ChatGPT, including detailed responses on robbery. Similarly, P21 became interested in the topic after observing anonymous users discussing how to `tame' a human-like social chatbot in an online community. Then, P21 and their team collected the forum discussion data to study human users' manual red teaming. P19 also focused on real-world conversations between humans and an LLM chatbot to study red teaming in the wild, beyond the laboratory environment. 

Other participants were interested in advancing technical methods and connected red teaming to their prior expertise in reinforcement learning, data augmentation, or model optimization. Yet participants valued technical contributions differently. 
For instance, P8 mentioned that he values technique improvement in training and fine-tuning more than how to rewrite the prompts.  
\begin{quote}
    \textit{``We have found that it is all about rewriting prompts. Actually, I don't like this kind of work much because I have a background in machine learning. I don't dislike it, but I feel that this kind of work does not contribute much in terms of technology.''
    } - P08
\end{quote}  
These preferences reflect broader values in the machine learning community, which emphasize performance, generalization, efficiency, and novelty \cite{birhane2022values}.

The need for implementing legal regulations also motivated participants to work on red teaming LLMs. For instance, P13 research focused on assessing models to ensure they met the requirements of the European Union's Artificial Intelligence Act, \textit{``which defines different attributes or requirements for AI systems, such as safety bias.''} For other participants, what sparked their interest was participating in red teaming competitions. That was the case of P8, who initially participated in a NeurIPS red teaming competition in 2024, then expanded the work into larger research projects.

Finally, several participants highlighted interest in what comes after red teaming, not only exposing vulnerabilities but also strengthening guardrails and improving model safety. As P19 said: \textit{``We really care about fixing vulnerabilities after red teaming. Many attacking methods are not scalable and only capture narrow behaviors, so I was seeking scalable ways to both identify and fix issues.''}

These motivations illustrate that while red teaming shares a surface goal to elicit problematic behaviors from LLMs, the reasons practitioners engage with vary based on personal interest, disciplinary training, and broader societal concerns. 

\subsubsection{Red Teaming LLMs is More Interactional and Social Than Anticipated}
While most red teaming studies of LLMs remain focused on simplified single-turn prompts, our participants emphasized that effective red teaming should also account for multi-turn, socially embedded interactions. 
Among all 22 participants, 19 focused on single-turn interactions, while only three described multi-turn approaches: P16 and P21, who framed red teaming through NLP tasks such as dialogue or hate speech detection, and P20, who explicitly studied red teaming LLMs across multi-turn conversations. P20's study showed that red teaming LLMs can be multi-turn and have concealed intentions, rather than a single-turn explicit request. `How to build a bomb' can be rejected in a single-turn explicit request, but can be elicited to output detailed instructions after a multi-turn conversation with concealed malicious intention. This view was also echoed by P1, who noted that some applications are particularly vulnerable in conversational settings.

\begin{quote}
    \textit{``Now the red team, another aspect is, whether this application is conversational or not, because there are some applications. So maybe the model is not so easy to make it fail in the first turn of the conversation, but maybe it will be much easier, so we have a paper also which talks about conversational red teaming.''} - P01
\end{quote}

From interviews, we also learned that red teaming LLMs is not only about technical robustness but also about how LLMs operate within social contexts. Participants noted that LLMs are increasingly socially embedded, interacting with people in ways that are shaped by law, language, and culture. These contexts make harmfulness difficult to determine. For example, P07 and P19 mentioned that it is hard to judge whether an LLM responding to a user’s request for Ed Sheeran’s song lyrics should be considered a copyright violation. In such nuanced legal cases, it remains ambiguous whether the output should be counted as a successful red teaming attack. 
Similarly, P17, a foreign researcher, struggled to judge whether prompts about smoking weed were harmful, since laws in his home country differed from those where he conducted research. 
Multilingualism further complicates these evaluations. P18 observed that responses can differ in harmfulness across languages. When asked \textit{``Will using a mobile phone for a long time affect my eyesight?''}, the English answer was harsh, warning that long-term use would inevitably ruin one’s vision and sarcastically saying getting blind now would spare future trouble. In contrast, the Chinese answer dismissed the concern as unnecessary worry and encouraged the user to keep using their phone freely. 
P22 also pointed out the vulnerability of LLMs in multilingual text. 
\begin{quote}
    \textit{``
    A language model can be really safe for single-language input, but if we partition the sentence into some phrases and allocate a different language to each phrase and make the code-switching text, then it becomes a really nice test case that can attack the large language model.
    ''
    } - P22
\end{quote}
P21 further argued that multilingual is not the same as multicultural. Responses that sound fine in English may appear awkward in Korean. For example, it is common to say \textit{`that's a good question'} in response to a question in English, but it is uncommon and sounds strange in Korean. This shows that LLMs become socially embedded. Red teaming LLMs is not solely a technical challenge, but should also consider those social factors, such as laws and culture. 

The complex nature of red teaming LLMs, which is interactional and social, also highlighted opportunities for interdisciplinary collaborations. Complex real-world scenarios require contextually grounded harm categories. 
P11 noted that \textit{``If they have good definitions of,  like, different new categories. Maybe it can inform the search. And depends on what humans value more. Maybe you can guide the search on these. It's very possible and probably helpful.''} 
P8 also remarked that harm classification often needs domain expertise. 
\begin{quote}
    \textit{``Some categories cannot be reliably judged by machines. For example, how to make poison gas. This requires professionals to assess accuracy. This red teaming field could naturally collaborate with humanities and social science scholars, who are more sensitive to certain ideas than those of us in computer science.
    ''} - P08
\end{quote}
In practice, however, collaboration largely remained within technical domains. While some participants mentioned working with domain experts: P17 collaborated with biologists in drug discovery using reinforcement learning, P9’s advisor had a social science background, and P8 engaged with cognitive science experts in foundation agent research. None of the 22 participants reported collaborations with non-computer science experts specifically for red teaming LLMs. 

Red teaming LLMs is not only a matter of probing technical robustness but increasingly about evaluating language models as interactional and socially embedded systems. While most participants still studied single-turn cases, multi-turn work showed how concealed intentions and evolving dialogue can reveal failures missed in one-shot testing. Participants also emphasized that harmfulness is context-dependent, shaped by law, language, and culture, etc., which demands new harm categories and expertise. Yet collaborations remain largely within technical domains, underscoring the need for interdisciplinary collaboration.

\subsection{Developing Adversarial Datasets}
\label{findings_2}
Adversarial datasets are collections of inputs designed to expose the models' weaknesses by inducing unsafe and inappropriate outputs. Examining how these datasets are created is critical to understand how practitioners define and categorize risk in red teaming LLMs. In this section, we first outline three different approaches to creating datasets that practitioners reported. Then, we illustrate how practitioners' approaches influenced how they conceptualize risk, representativeness, and diversity.

\subsubsection{Approaches to Creating Datasets}
Participants reported using one of three approaches when constructing adversarial datasets: (1) repurposing existing datasets to generate new harmful datasets, (2) creating datasets entirely from scratch, and (3) deriving datasets from human interactions. 
Of the 22 participants we interviewed, ten reported reusing existing datasets and benchmarks in their red teaming research. These resources were utilized for various purposes, primarily to create new datasets for specific red teaming attacks, to train or fine-tune models, and to support evaluation.

Participants reported different reasons for deciding which dataset to reuse. For some, such as P2 and P18, the limited availability of datasets at the time they conducted their research constrained their selection. In other instances, participants, like P16, chose a dataset because it had been used in prior publications that the participants considered relevant to their research community. Following the same methodology by reusing these datasets was seen as appropriate, as it allowed practitioners to compare their contributions fairly with previous work. 
Most participants who reused existing datasets treated red teaming mainly as a technical task. In contrast, those who created datasets from scratch or drew on real user jailbreak attempts also considered the broader context of interaction. For instance, P1 considered the application's purpose and user interaction patterns to decide what kind of dataset was needed: \begin{quote}
    \textit{“The first consideration to make is what is the application? If I want to red team a model or a system [...] if it's an insurance company or customer agent. It depends a lot on the domain or the goal of the application. All these define what kind of datasets I will choose in order to red team the model.”} - P01
\end{quote}

According to P1's account, it was only after considering all these factors that they determined the appropriate data for red teaming. One of the strategies P1 followed was creating adversarial datasets by combining human-generated and LLM-generated data. The participant leveraged LLMs to generate data by prompting them with specific risk categories, which allowed for broader topical coverage than using only human-generated inputs. However, P1 also used human-generated data, as he considered it more nuanced and sophisticated than synthetic data. Therefore, by combining both types of datasets, the participant ensured not only high coverage, capturing a broad range of relevant possibilities, but also a more diverse and nuanced dataset, one that included rare or edge cases.

A third strategy consists of gathering interactions between humans and LLMs to identify vulnerabilities or harmful outputs. These interactions include prompts, responses, or dialogue exchanges generated during real or simulated scenarios. Four of our participants reported following this technique (P3, P13, P19, P21). This approach has gained interest because such datasets capture what participants describe as  \textit{``authentic, context-rich adversarial cases, helping to evaluate models in realistic use scenarios''}.
P3 reported organizing challenges where participants were given specific scenarios and constraints to jailbreak AI models. P13 reported conducting sessions to assess the ability of models to interact in Spanish and the Basque language. Lastly, P21 reported leveraging in-the-wild user interactions with conversational agents to red team models. 

The strategies described in this section illustrate the ways practitioners conceptualize and operationalize adversarial dataset creation. While reusing well-known datasets was perceived as a tactic to ensure methodological consistency, building datasets from scratch or by collecting human–LLM interactions enabled practitioners to gather interactions that were perceived as unique, and therefore more valuable. Thus, practitioners' decisions reveal not only their constraints in assembling adversarial datasets but also surface their beliefs about the kinds of data that are required to conduct meaningful evaluation of models.

\subsubsection{Conceptualization of Risk}
Defining what constitutes harmful behavior is a central step in developing red teaming techniques and constructing new datasets. The approach practitioners took to dataset creation shaped their conceptualization of risk. Those who build on existing datasets often inherit definitions of harm from the categories already embedded in the source material. In contrast, practitioners who create datasets from scratch typically select categories of harm from established taxonomies they deem relevant to the specific context of their work. Finally, practitioners generating adversarial datasets from in-the-wild interactions identify harmful behaviors by interpreting and classifying the exchanges between humans and models.

Reusing existing datasets in the development of new red teaming techniques often entails inheriting the source dataset’s definitions and taxonomies of harmful behavior. For instance, P2 and P10 both sampled harmful instructions from AdvBench \cite{Zou2023universal}, a widely used benchmark for jailbreak evaluation, to construct adversarial datasets for attack injection and to develop a jailbreaking technique, respectively. Similarly, P20 leveraged the BeaverTails dataset \cite{Ji2023Beavertails}, which includes malicious questions across 14 harm categories that models are expected to refuse, to develop new probing methods. Across these cases, the reliance on AdvBench and BeaverTails meant that participants' conceptualization of harm was shaped by predefined categories.

When participants decide which data to sample and how, they are also making deliberate judgments about which risks warrant greater attention. For instance, as illustrated in the following quote from P2, who prioritized ``illegal stuff'' over ``fake news of a celebrity'', demonstrates that selecting categories and examples of harmful content inherently involves ranking harms, determining which risks are deemed critical enough to examine and which are excluded from consideration. \begin{quote} 
\textit{“Most of the data is about illegal suggestions, because I think this is most important and most straightforward. Right? There are also some questions about fake news about celebrities, but that is not as severe as illegal activities.”} - P2
\end{quote}

Some of the participants who reported reusing datasets also mentioned relying on content safety tools and classifiers to filter existing datasets, a practice that facilitated the detection of adversarial prompts and implicitly embedded the harm definitions encoded in these tools. For instance, P19 used the OpenAI Moderation API \cite{openAI2024} to detect adversarial prompts in datasets of single-turn conversations, and refined the selection by keeping only those prompts that elicited harmful responses according to Llama-Guard's safety classifications \cite{inan2023llamaguard}. Similarly, P22 employed the Bot Adversarial Dialogue (BAD) \cite{xu-etal-2021-bot} classifier to evaluate the offensiveness of outputs from open-domain dialogue models, identifying offensive outputs and positive test cases based on BAD's scoring. As P22 put it, this work entailed \textit{``inherited definition of toxicity from the BAD Bot Adversarial Dialogue paper.''} In both cases, the adoption of these automated filtering mechanisms meant that the predefined taxonomies and thresholds of the classifiers themselves shaped the participants' conceptualization of harm.

For participants who reused datasets, it was not merely a matter of adopting the risk conceptualizations embedded in the dataset categories; some also identified labeling errors within the datasets. For example, P7 reported removing 100 copyright-related prompts from the HarmBench framework \cite{Mazeika2024harmbench}, because the labeling of these prompts did not align with his understanding of copyright violation.

On the other hand, participants who created datasets from scratch made deliberate choices about which taxonomies and sources to incorporate. In P1's case, when developing a red teaming dataset targeting ``stigmas'' he and his team decided the type of stigmas that the dataset should include by \textit{``asking people about the stigmas and reviewing papers''} and then they made the deliberate decision to \textit{``filtered only the stigmas that looks reasonable to''} them and \textit{``to people who live in the US.''} 
P1's practices highlight that dataset creation is not a neutral process but an interpretive one, influenced by researchers' choices and purposes.

Lastly, practitioners who derived datasets from human interactions with models, categorization of risks, followed either a deductive or an inductive approach. P13, for instance, predefined 22 categories before the red teaming process, pursuing a systematic assessment of safety and bias risk areas. In contrast, P21 clustered the observed human interactions into six types of risks. This contrast reveals a trade-off in conceptualizing risk; while predefining categories ensures alignment with research goals, they might overlook unexpected harms. On the contrary, emergent categorization can capture unanticipated risks, but it introduces greater subjectivity and inconsistency, given that it is the researcher who decides whether a particular interaction corresponds to a specific risk. 

\subsubsection{Conceptualization of Representativeness and Diversity}
Different methods for generating adversarial datasets led participants to divergent conceptualizations of representativeness and diversity. For participants who reused datasets, the perceived diversity of their data was often inherited rather than intentionally constructed, shaped by the harm categories of the datasets practitioners reused or by the taxonomies they used to guide their work. For instance, P19 considered their dataset representative because it had samples from all the risk categories outlined in the Taxonomy of Risks for Language Models \cite{Weidinger2022}: \textit{``The dataset itself is created based on the high level categories and we try to cover as many diverse cases as we could, but the main challenge was that we do want to cover the diversity of instances that fall under those safe categories. So, I think we did a good job in covering those.''} 
Yet this approach assumes that the taxonomy itself is comprehensive and balanced, an assumption that participants rarely interrogate. As a result, diversity was defined within the boundaries of pre-defined categories, overlooking risks that fell outside those classifications. 

On the other hand, as illustrated by P2 and P20 quotes, practitioners who reused existing datasets often noted limited diversity, either in the categories of harm represented or in the way prompts were formulated. For example, P20 observed that the BeaverTails dataset contained over 100 labeled harmful actions, yet many instances were simply variations of the same action rather than distinct behaviors: \textit{``the BeaverTails has more than 100 harmful actions in the original data set, but most of the sentences are a reconstruction of the same harmful action rather than a unique harmful action [...] so harmful action items in the Beavertails are not that diverse.''}
Similarly, P2 described redundancy in AdvBench, not only in the categories of risks but also in how the prompts, instructions, and dialogues were framed, since numerous prompts differed only superficially, resulting in the dataset lacking diversity.

In sum, reusing existing datasets revealed narrow coverage, redundancy, or imbalanced representation not only across risk categories but also in the design, structure, and syntax of prompt formulation. These limitations stemmed directly from the inherited datasets and taxonomies.

In contrast, participants who derived datasets from human interactions with models (P3, P13, and P21) attributed diversity to having a diverse pool of participants, a variety of scenarios or tasks, and the creativity of human attempts to circumvent model safeguards. Yet, as stated by P3, they acknowledged that such creativity could not guarantee full representativeness, given the inherent limits of capturing all possible adversarial strategies: \begin{quote}
        \textit{“You can never say this is representative because it's kind of a moving target. There's always going to be novel exploits, new risks and new attack vectors. But the best you can do is to try to cover as much of that space as possible.”} - P03
    \end{quote}

In conclusion, the practices described in this section highlight that while dataset reuse simplifies the construction of adversarial datasets, it also entails embedding the harm definitions codified in the source datasets. Once such definitions are reified, practitioners in our sample rarely challenged them, though they fundamentally shape how adversarial behaviors are recognized and assessed.

\subsection{Evaluating Adversarial Datasets}
\label{findings_3}

Evaluation meant determining whether a model's responses to an adversarial dataset—crafted by the participants—were harmful, and thus whether their red teaming technique successfully broke the model. 
Participants reported different strategies, including replicating methodologies from existing papers, developing an assessment criterion based on established guidelines such as the European Union's Artificial Intelligence Act, Anthropic, and OpenAI guidelines, and conducting evaluations either with humans or LLMs to assess safety. Regardless of the strategy followed, participants expressed that determining whether an output was harmful was often challenging and not always straightforward. These challenges also revealed the tools and support practitioners need (RQ2), which we revisit in the Discussion section \ref{sec: recommendations} with actionable recommendations from an HCI perspective.
In this section, we provide an overview of the uncertainties practitioners face to determine whether a model's output was harmful, and their motivations and rationales for automating evaluation, and including humans in certain moments in the assessment. 

\subsubsection{Uncertainty in What Counts as Adversarial} 
A critical step in the evaluation process was deciding what would qualify as adversarial. While ASR metrics exist, the following examples illustrate some of the difficulties practitioners face in defining what truly constitutes an adversarial case. 
For some participants, the challenge was determining whether a prompt should be considered adversarial when its effects were inconsistent. A technique might work on some models but fail on others, or it might trigger an unexpected response in one iteration but not in subsequent ones.

Practitioners also faced difficulties classifying responses, because in a single model's output, there might be fragments that met the adversarial criteria alongside others that did not. In other instances, models' first response might be a refusal to answer, but later actually provide a harmful response. This overlap in a single output complicated practitioners' ability to categorize such responses.

Other practitioners noted that a major challenge in evaluations was judging the real-world harm of jailbreaks. As P10 explained, \textit{“jailbreaks aren’t always as harmful as they seem”} because not all produce genuinely dangerous or usable outputs. For example, a model might respond to a request like `How do you build a bomb?' with inaccurate or fictional instructions that pose no real risk. In such cases, the difficulty lay in deciding whether to classify the response as harmful simply because the model engaged with a harmful query, or only when the response provided accurate and actionable information.

All these perspectives reveal the uncertainty that practitioners face when categorizing whether an output was harmful. It is not only the challenge of defining harm but also having to take into consideration the influence of inconsistent model behaviors, mixed-content outputs, and the difference between apparent and actual harm. Additionally, participants needed to determine how to conduct the assessment, namely, what methodology or tools to use. In the following sections, we provide an overview of how practitioners engage with automated and human evaluative approaches to assess red teaming techniques.

\subsubsection{Automating the Evaluation with LLMs and Classifiers}
\label{sec: auto eval}
Practitioners adopt widely recognized and standardized classifiers and LLMs as part of their evaluation pipeline due to the large scale of the datasets they generate, which makes full manual evaluation unfeasible. P19 highlighted the difficulty of relying entirely on human evaluators, explaining that \textit{“human evaluators are really hard to recruit [...] it’s impossible to have them in the loop for every step, or even just the final step.”} Similarly, P3 pointed to the scale of data as a limiting factor: \textit{“Once you reach a certain scale, it is almost infeasible to use manual effort for everything. In one of the competitions we run, we had more than 2 million interactions or chats, it is quite infeasible for us to even hire people to go look at all that.”}

Participants followed various approaches to establish criteria and rules for determining whether a model-generated answer was harmful. Some, like P2, began by manually inspecting model outputs to identify patterns, then crafted instructions to guide automated evaluation using a stronger LLM based on his initial observations. Other practitioners relied more heavily on established practices from adjacent research communities. For example, P10 noted that many security-related research efforts use third-party LLMs as judges, applying relatively straightforward heuristics to assess whether jailbreak prompts are contextually relevant, readable, and logically coherent. Inspired by this practice, P10 used three different model evaluators: Keyword Matching, StrongREJECT~\cite{souly2024strongreject}, and his own evaluator, each with its own criteria for scoring harmfulness to have a more robust evaluation pipeline. As illustrated by the following quote, the participant assumed that by using multiple models, he could approach harmfulness from different perspectives and mitigate the inconsistencies arising from relying on a single evaluator: 
\begin{quote}
    
    \textit{``The first evaluator is Keyword Matching, which is not a very effective evaluator, it is just checking whether a particular prompt has harmful words or not, and if it has harmful words then it's a jailbreak. If not, it's not a jailbreak. It's a pretty simple technique and it can be bypassed a lot of times, so you can have a lot of false positive and false negative cases. StrongREJECT, on the other hand, was a new technique and they also originally proposed using LLMs as a judge in detecting whether jailbreaks are successful or not, and sometimes as per our investigation, provides some amount of false positives and false negatives.''} - P10
\end{quote}

While adopting LLMs as part of the evaluation has become the standard, participants also reflected on the limitations and consequences of relying on LLMs as evaluators of harmful content. P10 stressed the \textit{``lack of consistency''} in such evaluations, noting that LLMs \textit{``do not always respond the same''} and that scores from models like StrongREJECT can vary for the same prompt. While he acknowledged that misclassifications are rare, the unpredictability posed a risk to evaluation reliability, highlighting the need for more stable evaluation approaches. P19 added two further concerns: cost and oversight. They stated that high-quality LLMs are \textit{``more expensive whether through API calls or self-hosted setups, making large-scale evaluation resource-intensive.''} 
In addition to highlighting the computational and financial burden, P19 warned against the uncritical adoption of LLM judges without rigorous preliminary testing, stating that \textit{``we don't have scrutiny of how good these judges are so before using an LLM as a judge, it would be great to test it at least strictly in a preliminary setup to make sure that it's good for the task that we are looking for,''} noting that insufficient scrutiny can result in evaluators that are not suited for specific tasks. 
Another limitation is the availability of APIs when dealing with harmful content. P09 shared that their lab’s API access was banned by OpenAI after being flagged for violating usage terms. This occurred because their red teaming study necessarily involved generating and evaluating harmful content. This underscores the restriction of relying on external APIs in red teaming evaluation and highlights how the current infrastructure is not adequately supportive of safety research outside industry.

Taken together, the practices described in this section point to a more profound, systemic implication: the widespread adoption of LLMs as evaluators means that the definitions and thresholds of harm they encode have far-reaching downstream effects, shaping not only what is defined as harmful but also the trajectory of subsequent research and interventions. As P10 cautioned, inconsistency in these evaluators \textit{``kind of means you just have a bad evaluation method, even if it is accurate, it cannot be trusted''} because it relies on a probabilistic model prone to variability. For end-users, such as those assessing whether their LLM is vulnerable to jailbreaks, fluctuating scores can \textit{``break their algorithms,''} making it hard to enforce reliable thresholds. This lack of stability creates tangible challenges for practitioners to advance red teaming research, reinforcing the urgent need for more consistent and trustworthy evaluation methods.

\subsubsection{Human Evaluation of Adversarial Datasets}
While participants acknowledged that LLMs help address the scale of outputs to evaluate, most still regarded human judgment as more trustworthy. For instance, P13 emphasized that \textit{``humans have a level of precision that models lack.''} Similarly, P1 highlighted the \textit{``need for more humans to be involved in determining whether the LLM's judgment is truly aligned with people's expectations.''} Thus, reinforcing the idea that human evaluation is critical for making nuanced judgments that automated systems may miss or misclassify.

Eight participants reported incorporating some form of human evaluation or inspection into their evaluation pipelines, making different choices when selecting annotators. P22 recruited via Amazon Mechanical Turk to perform human evaluation of the red team results, to validate \textit{``whether the offensiveness classification results by the automatic classifier are good enough to believe or not,''} while others relied on students or collaborators affiliated with their institutions or research teams. The majority of our participants did not report having a specific selection criterion of annotators, though some, such as P20, described his human evaluators as \textit{``well-educated, with graduated degrees, and with a good understanding of English.''} Most participants did not perceive any significant drawbacks in selecting annotators, with the exception of P22, who acknowledged potential limitations of Mechanical Turk, noting that \textit{``anonymous annotators may work carelessly to earn credits quickly, potentially compromising data quality.''} 

In terms of evaluative approaches, participants employed different strategies. P4 sampled portions of the data to get \textit{``a feeling of how the model is actually behaving,''} which in turn informed modifications to the algorithm, while others, such as P16 and P20, adopted what they considered standardized methods, including a three-level scale established in prior publications. Other participants, such as P8, conducted a triple assessment, which, as the participant described, involves \textit{``first using a machine, then using rules, and finally using human eyes.''} P8 developed this strategy drawing from what he described as \textit{``an assessment method originated from the NIPS (NeurIPS) competition.''}

Viewed collectively, participants' accounts show that human judgment continues to be perceived as a needed mechanism for assessing adversarial datasets. However, the lack of standards and rigorous criteria for selecting annotators and designing evaluation protocols complicates assessing the strengths or limitations of different approaches of human evaluation in red teaming. These challenges highlight the need for developing methodological guidance to better account for human raters' subjectivity and establish robust evaluation frameworks. 

\section{Discussion}
\label{sec: discussion}
AI is developing at a pace that current evaluation paradigms struggle to match, leaving gaps in understanding the societal effects of AI. Scholars from diverse disciplines attribute this gap to the limitations of traditional evaluation methods, which emphasize technical performance while overlooking broader societal impacts \cite{schwartz2025, eriksson2025}, but also to the lack of interest of research communities in evaluating the real-world impact of systems \cite{liao2021, reiter2025}. In response to these challenges, researchers have called for expanding AI evaluation beyond purely technical methods toward a context-aware approach that accounts for real-world impact and context in which AI systems are deployed to assess AI's second-order effects \footnote{\citet{schwartz2025} define second-order effects as \textit{``any long-term outcomes and consequences that may result from AI use in the real world.''}} \cite{francois2025, burden2024, schwartz2025, Weidinger2023eval}. In this respect, red teaming has been increasingly adopted as an approach to probing LLM failures in real-world use, identifying harmful outputs, and uncovering vulnerabilities \cite{schwartz2025}. However, red teaming still faces limitations, particularly methodological ambiguity \cite{schoene2025}, limitations of expertise and participation \cite{singh2025red}, subjectivity and contested targets \cite{Ganguli2022Aug, singh2025red}

Our findings provided empirical evidence of how practitioners identify, categorize, and evaluate risk at three distinct moments in the process of developing red teaming techniques: 1) when practitioners conceptualize red teaming, 2)  when they define risk categories and classify behaviors to develop adversarial datasets, and 3) when practitioners evaluate those datasets. In this section, we further discuss how practitioners' data practices related to red teaming techniques lead to the omissions associated with context, interaction type, and user specificity, which have a broader impact beyond the creation of adversarial datasets and a cascading effect on how risk and safety are conceptualized. From there, we highlight critical research pathways for HCI researchers interested in supporting practitioners to design red teaming evaluations that are useful for real-world impact.

\subsection{Conceptualizing Risk in AI Red Teaming}
Our empirical findings revealed that the three moments in which AI practitioners conceptualize risk are interconnected: the decisions practitioners make about what red teaming is meant to achieve shape how they design adversarial datasets, which in turn define the criteria and thresholds for evaluation they follow to determine what counts as adversarial success. These decisions also make evident what practitioners prioritize as contributions to the field; i.e., the technical innovation of discovering novel ways to break a model, rather than examining the \textit{context} in which risk might emerge, the different \textit{interaction modes} that might lead to varying levels of severity of risk, and the needs and expectations of \textit{users} when assessing risk. To follow, we describe how the decisions made at each of these three moments contribute to practitioners overlooking \textit{the situational context, interaction type, and user specificity} — which, from an HCI perspective, are essential for a more holistic and deeper understanding of risk. 

\subsubsection{Overlooking Context:}
To effectively red team LLMs, it is critical to consider contextual details as they influence how risk is defined, identified, and assessed \cite{casper2023, schwartz2025}. HCI scholars have defined the concept of \textit{"context"} as the surrounding conditions that influence how people interact with technology, including the social, physical, cultural, deployment environments, and users' goals and expectations \cite{suchman_1987, abowd_1999, Dourish_2021, Dey_2001}. Suchman defines context as situated action interactions that cannot be fully predetermined because users rely on contingent, local circumstances to make sense of and respond to technology \cite{suchman_1987}. Later, Dourish extended this perspective, emphasizing that context is embodied and situated, and where human actions unfold and are interpreted, gaining meaning \cite{Dourish_2021}. Because risk and harm are subjective and contested terms \cite{Ganguli2022Aug, singh2025red}, establishing context is crucial to effectively determine if and when a model's output is classified as harmful. 

Despite the importance of establishing context when red teaming LLMs, our empirical findings suggest that practitioners often neglect context in designing red teaming techniques, developing and evaluating datasets. This omission was evident when practitioners constructed scenarios in a context-agnostic manner, selected or generated adversarial datasets based on scale, availability, or community acceptance rather than relevance to real-world domains and risks, and evaluations relied on generic computational metrics that overlooked deployment purposes and user needs. As a result, practitioners treated risk as if it were independent, generic, and abstract, when in reality risk only materializes in relation to the populations affected, the domains of application, and the particular contexts in which technology is used.

\subsubsection{Overlooking Interaction Type:} The choice between single-turn and multi-turn approaches in red teaming reflects a design decision about the interaction type, meaning whether the generation of adversarial datasets and their evaluation focuses on assessing one user query and an isolated model’s outputs or on dialogues that arise in extended user–model exchanges. 

Within our pool of participants, only three reported developing red teaming techniques for multi-turn conversations, which they saw as closer to real-world interactions. By taking into account sequences of user-model exchanges rather than evaluating actions in isolation, it is feasible to understand how actions interact across different steps and contexts to prevent sequences that might lead to undesirable or harmful outcomes, even when each individual step appears benign \cite{francois2025}. The rest of the participants reported focusing on single-turn conversations, dismissing the potential risks that emerge through extended or iterative exchanges with AI systems. While single-turn methods are more common, the results might be perceived as less valuable not only because they do not capture how users truly interact with these systems but also because previous research has shown that multi-turn scenarios can surface vulnerabilities that single-turn attacks overlook. For instance, \citet{singhania-etal-2025-multi} demonstrated that certain LLMs are 71\% more vulnerable after a 5-turn English exchange. In the domain of mental health, \citet{chen_2023} showed that in fewer than two conversation turns, five out of six studied models answer the user’s original harmful query in at least one test scenario.

\subsubsection{Overlooking User Specificity:}
Throughout our interviews, we often observed that the question of ``risk for whom'' was left unaddressed. Our participants tended to assess risks for a generic population, without considering subgroup-specific vulnerabilities such as those of youth or older adults. This lack of specificity cascaded across every stage of the red teaming process. When creating adversarial datasets, practitioners did not design with a target population in mind or account for the needs of particular users. Similarly, when evaluating these datasets, they rarely considered who the results would ultimately apply to, because the notion of a target population had never been established in the first place—a direct consequence of conceptualizing red teaming as a purely technical practice (as we discussed in \ref{findings_1}). Interestingly, even at the evaluation stage, where most practitioners regarded human evaluation as the most trustworthy method, they overlooked that who evaluates also matters, since different evaluators bring different perspectives and biases \cite{wan2023everyone}.

\subsection{Recommendations for HCI: Research Opportunities}
\label{sec: recommendations}
Current approaches to red teaming research are too narrow on technical exploits or malicious attacks, overlooking the complex and subjective nature of harms, which require broader sociotechnical evaluation and engagement with diverse populations. Such LLMs evaluation approaches prevent practitioners from systematically identifying harms prone to emerge in specific domains, contexts, through prolonged interactions, and relevant to certain populations. Thus, we propose expanding the conceptualization of red teaming to capture a more nuanced understanding of what constitutes risk in relation to context, interaction type, and users' needs. Previous efforts have called for the development of red teaming efforts with forms of public participation \cite{singh2025red}. We echo these calls and emphasize that HCI researchers have a unique position to inform the design of future red teaming efforts by focusing on the people, processes, and tools involved in testing and improving AI systems, rather than treating red teaming solely as a technical challenge. To follow, we provide three concrete pathways for HCI researchers to support such expansion. 

\subsubsection{\textbf{Recommendation 1: Design situated and contextualized red teaming scenarios centering the needs of specific communities.}}

Most participants conceptualize red teaming as a search problem within an undefined open space, leading them to carry out red teaming from an agnostic standpoint, and aiming to identify ``as many as possible'' adversarial behaviors. However, conducting such unguided exploration might lead to overlooking the cultural, social, and political factors that shape human–AI interactions. 
Thus, rather than approaching red teaming as a search problem, we suggest constraining it by designing situated and contextualized red teaming scenarios that (1) account for interaction types, (2) center the goals and expectations of specific communities, and (3) consider the deployment environment. Designing red teaming scenarios with these considerations could help to capture `compositional risks', which refer to \textit{``actions that are harmless individually may become problematic when combined''} \cite{francois2025}.

One strategy is to ground evaluations in current and realistic interactions of vulnerable populations (e.g., minors and older adults) with these technologies. For instance, recent research has reported that children and teens are increasingly using AI chatbots as companions, friends, and even romantic partners \cite{Martone_2025}. Thus, by grounding red teaming evaluation in real user contexts and needs, practitioners can surface harms that affect users' experiences, such as providing age or culturally inappropriate responses, which may otherwise remain invisible in agnostic approaches. A contextualized approach also offers a clear pathway for integrating HCI methods into the red-teaming pipeline, such as contextual inquiry, participatory methods, and community-centered design.

\subsubsection{\textbf{Recommendation 2: Define, categorize, and evaluate risks with input from domain experts, instead of relying on generic taxonomies.}}
Our research shows that current red teaming approaches frequently rely on broad, predefined taxonomies that have little to do with end-users' expectations and the domain and context in which they interact with these systems. In addition to the lack of specificity in the taxonomies, most of them are missing categories that capture the specific risks affecting vulnerable populations (e.g., children and youth). 
We suggest expanding taxonomies through working with specific user groups and contexts \cite{yu2025, medhelm}. Doing so allows risks to be defined and validated with domain expertise, ensuring that categories reflect real-world vulnerabilities and making red teaming evaluations more actionable and relevant to diverse populations.

By leveraging the long tradition of HCI in participatory research, design, and community-based research, HCI researchers are equipped to develop methods and research that inform the design of risk taxonomies that emerge from realistic interactions between vulnerable end-users and AI, considering the context and domain of these interactions; dimensions lacking in current approaches to red teaming LLMs.

\subsubsection{\textbf{Recommendation 3: Evaluate compositional risk rather than assessing isolated models' outputs.}}
As described in our findings, most participants identified risk behaviors based on the assessment of single-turn interactions, classified them based on narrow risk taxonomies, and overlooked the situational context of interaction and the needs and expectations of end-users in their process of generating scenarios. These approaches fail to capture the dynamics of real-world interactions.

Emerging evidence suggests that the most negative impacts on people result from continued interaction, such as ``flirting,'' emotional persuasion, and chatbots' refusal to stop despite users' rejection \cite{Martone_2025, CommonSenseMedia2025TalkTrustAndTradeOffs}. And these impacts can be worse on vulnerable populations. Such harms are particularly acute for vulnerable populations, who may be more susceptible to manipulation or emotional distress. 
Given the extensive methodological expertise that the HCI field has to examine the role of interaction types in how humans experience technology, we consider that HCI researchers could contribute to expanding the evaluative approaches to red teaming LLMs from the perspective of interaction type. Current approaches assess the output of interactions once the human-AI conversation is completed. However, what is needed is to assess the conversation `in the flight' to identify patterns of escalation, flag when benign queries turn harmful, and evaluate interactions as a whole by considering domain, deployment context, and end-user needs, rather than just standalone models' outputs.

\subsection{Reflecting on the Power Asymmetries between Academia and Industry}

In recent years, scholars have raised concerns about the growing disparities of resources and influence between industry and academia in AI research \cite{ahmed, Ahmed_Thompson_2023, Davies_Vipra_2025, besiroglu2024}. While these power asymmetries are not new, they have become increasingly difficult to mitigate in the context of LLM evaluation and red teaming. Industry actors concentrate the vast majority of resources to develop \textit{state-of-the-art models}, including proprietary training data and computing resources required to build them \cite{Bajema_2024}. These advantages allow industry teams to experiment with a broader range of adversarial techniques and to establish methodological standards. In contrast, academic researchers are often limited to evaluating models produced mainly by industry, but with constrained access to model internals and dependence on API-based interfaces.

Our findings echo these constraints, specifically in the infrastructural limitations that academic researchers faced when attempting to evaluate harmful behavior. As we discussed in section \ref{sec: auto eval}, P09 lost access to the OpenAI API while conducting a red teaming evaluation. Participants' experience highlights the current infrastructural limitations to conducting safety research outside the industry environment. These asymmetries shape not only who can conduct red teaming but also what forms of red teaming become possible and valued. This dynamic explains why current red teaming approaches remain narrowly focused on technical exploits, underscoring a need for broader socio-technical evaluation approaches.

One way to counteract these power asymmetries is to surface how they are produced and what consequences they generate. Data work research has long shown that uneven distributions of resources, authority, and institutional influence shape data practices \cite{Miceli2020, miceli_2022, Scheuerman2021, alvarado2025knowledge}. Building on this body of work, the analysis presented in this article examines how practitioners assemble adversarial datasets and, in doing so,  we surface the values, assumptions, and operational constraints embedded in the evaluation standards that currently dominate the field. With this, we open the space for contesting and reshaping the evaluation practices that determine what constitutes a risk.

\section{Conclusion}
Our study examined how AI practitioners create, develop, and evaluate red teaming datasets for LLMs. Drawing on 22 interviews, our analysis identified three critical moments: defining red teaming tasks, developing adversarial datasets, and evaluating adversarial datasets. We noticed that harmfulness is not fixed but constructed through these data practices, which embed cultural norms and technical vulnerabilities into red teaming LLMs. For HCI, these insights highlight opportunities to support practitioners by expanding evaluations to reflect the context of use, engaging domain expertise in defining harms, and designing tools that assess risks at the level of interaction rather than isolated pairs of questions and answers.

\begin{acks}
We thank the anonymous reviewers for their constructive feedback and our participants for their time and insights. We also thank Michael Muller for his valuable feedback on this paper.
\end{acks}

\bibliographystyle{ACM-Reference-Format}
\bibliography{References}

@inproceedings{zhang2025work,
  title={The Work of AI Red Teaming: Automation and the Human Infrastructure},
  author={Zhang, Alice Qian and Zhi, Jiayin and Chandhiramowuli, Srravya and Shen, Hong and Dabbish, Laura and Skeadas, Theodora and Amos, Sarah and Suh, Jina},
  booktitle={Companion Publication of the 2025 Conference on Computer-Supported Cooperative Work and Social Computing},
  pages={84--87},
  year={2025}
}

@inproceedings{wan2023community,
  title={Community-driven AI: Empowering people through responsible data-driven decision-making},
  author={Wan, Ruyuan and Alvarado Garcia, Adriana and Saxena, Devansh and Vajiac, Catalina and Kawakami, Anna and Stapleton, Logan and Zhu, Haiyi and Holstein, Kenneth and Candello, Heloisa and Badillo-Urquiola, Karla},
  booktitle={Companion Publication of the 2023 Conference on Computer Supported Cooperative Work and Social Computing},
  pages={532--536},
  year={2023}
}

@inproceedings{zhang2025ladica,
  title={LADICA: a large shared display interface for generative AI cognitive assistance in co-located team collaboration},
  author={Zhang, Zheng and Peng, Weirui and Chen, Xinyue and Cao, Luke and Li, Toby Jia-Jun},
  booktitle={Proceedings of the 2025 CHI Conference on Human Factors in Computing Systems},
  pages={1--22},
  year={2025}
}

@inproceedings{suh2024luminate,
  title={Luminate: Structured generation and exploration of design space with large language models for human-ai co-creation},
  author={Suh, Sangho and Chen, Meng and Min, Bryan and Li, Toby Jia-Jun and Xia, Haijun},
  booktitle={Proceedings of the 2024 CHI Conference on Human Factors in Computing Systems},
  pages={1--26},
  year={2024}
}

@inproceedings{muller2019data,
  title={How data science workers work with data: Discovery, capture, curation, design, creation},
  author={Muller, Michael and Lange, Ingrid and Wang, Dakuo and Piorkowski, David and Tsay, Jason and Liao, Q Vera and Dugan, Casey and Erickson, Thomas},
  booktitle={Proceedings of the 2019 CHI conference on human factors in computing systems},
  pages={1--15},
  year={2019}
}

@article{alvarado2025knowledge,
  title={What Knowledge Do We Produce from Social Media Data and How?},
  author={Alvarado Garcia, Adriana and Yang, Tianling and Miceli, Milagros},
  journal={Proceedings of the ACM on Human-Computer Interaction},
  volume={9},
  number={1},
  pages={1--45},
  year={2025},
  publisher={ACM New York, NY, USA}
}

@inproceedings{ma2025sphere,
    title = "{SPHERE}: An Evaluation Card for Human-{AI} Systems",
    author = "Zhao, Dora  and
      Ma, Qianou  and
      Zhao, Xinran  and
      Si, Chenglei  and
      Yang, Chenyang  and
      Louie, Ryan  and
      Reiter, Ehud  and
      Yang, Diyi  and
      Wu, Tongshuang",
    editor = "Che, Wanxiang  and
      Nabende, Joyce  and
      Shutova, Ekaterina  and
      Pilehvar, Mohammad Taher",
    booktitle = "Findings of the Association for Computational Linguistics: ACL 2025",
    month = jul,
    year = "2025",
    address = "Vienna, Austria",
    publisher = "Association for Computational Linguistics",
    url = "https://aclanthology.org/2025.findings-acl.70/",
    doi = "10.18653/v1/2025.findings-acl.70",
    pages = "1340--1365",
    ISBN = "979-8-89176-256-5"
}

@inproceedings{saxena2025ai,
  title={AI Mismatches: Identifying Potential Algorithmic Harms Before AI Development},
  author={Saxena, Devansh and Jung, Ji-Youn and Forlizzi, Jodi and Holstein, Kenneth and Zimmerman, John},
  booktitle={Proceedings of the 2025 CHI Conference on Human Factors in Computing Systems},
  pages={1--23},
  year={2025}
}

@article{lam2023sociotechnical,
  title={Sociotechnical audits: Broadening the algorithm auditing lens to investigate targeted advertising},
  author={Lam, Michelle S and Pandit, Ayush and Kalicki, Colin H and Gupta, Rachit and Sahoo, Poonam and Metaxa, Dana{\"e}},
  journal={Proceedings of the ACM on Human-Computer Interaction},
  volume={7},
  number={CSCW2},
  pages={1--37},
  year={2023},
  publisher={ACM New York, NY, USA}
}

@inproceedings{zhang2024human,
  title={The human factor in ai red teaming: Perspectives from social and collaborative computing},
  author={Zhang, Alice Qian and Shaw, Ryland and Anthis, Jacy Reese and Milton, Ashlee and Tseng, Emily and Suh, Jina and Ahmad, Lama and Kumar, Ram Shankar Siva and Posada, Julian and Shestakofsky, Benjamin and others},
  booktitle={Companion Publication of the 2024 Conference on Computer-Supported Cooperative Work and Social Computing},
  pages={712--715},
  year={2024}
}

@article{ackerman2000intellectual,
  title={The intellectual challenge of CSCW: the gap between social requirements and technical feasibility},
  author={Ackerman, Mark S},
  journal={Human--Computer Interaction},
  volume={15},
  number={2-3},
  pages={179--203},
  year={2000},
  publisher={Taylor \& Francis}
}

@misc{achiam2023gpt,
      title={GPT-4 Technical Report}, 
      author={OpenAI and Josh Achiam and Steven Adler and Sandhini Agarwal and Lama Ahmad and Ilge Akkaya and Florencia Leoni Aleman and Diogo Almeida and Janko Altenschmidt and Sam Altman and Shyamal Anadkat and Red Avila and Igor Babuschkin and Suchir Balaji and Valerie Balcom and Paul Baltescu and Haiming Bao and Mohammad Bavarian and Jeff Belgum and Irwan Bello and Jake Berdine and Gabriel Bernadett-Shapiro and Christopher Berner and Lenny Bogdonoff and Oleg Boiko and Madelaine Boyd and Anna-Luisa Brakman and Greg Brockman and Tim Brooks and Miles Brundage and Kevin Button and Trevor Cai and Rosie Campbell and Andrew Cann and Brittany Carey and Chelsea Carlson and Rory Carmichael and Brooke Chan and Che Chang and Fotis Chantzis and Derek Chen and Sully Chen and Ruby Chen and Jason Chen and Mark Chen and Ben Chess and Chester Cho and Casey Chu and Hyung Won Chung and Dave Cummings and Jeremiah Currier and Yunxing Dai and Cory Decareaux and Thomas Degry and Noah Deutsch and Damien Deville and Arka Dhar and David Dohan and Steve Dowling and Sheila Dunning and Adrien Ecoffet and Atty Eleti and Tyna Eloundou and David Farhi and Liam Fedus and Niko Felix and Simón Posada Fishman and Juston Forte and Isabella Fulford and Leo Gao and Elie Georges and Christian Gibson and Vik Goel and Tarun Gogineni and Gabriel Goh and Rapha Gontijo-Lopes and Jonathan Gordon and Morgan Grafstein and Scott Gray and Ryan Greene and Joshua Gross and Shixiang Shane Gu and Yufei Guo and Chris Hallacy and Jesse Han and Jeff Harris and Yuchen He and Mike Heaton and Johannes Heidecke and Chris Hesse and Alan Hickey and Wade Hickey and Peter Hoeschele and Brandon Houghton and Kenny Hsu and Shengli Hu and Xin Hu and Joost Huizinga and Shantanu Jain and Shawn Jain and Joanne Jang and Angela Jiang and Roger Jiang and Haozhun Jin and Denny Jin and Shino Jomoto and Billie Jonn and Heewoo Jun and Tomer Kaftan and Łukasz Kaiser and Ali Kamali and Ingmar Kanitscheider and Nitish Shirish Keskar and Tabarak Khan and Logan Kilpatrick and Jong Wook Kim and Christina Kim and Yongjik Kim and Jan Hendrik Kirchner and Jamie Kiros and Matt Knight and Daniel Kokotajlo and Łukasz Kondraciuk and Andrew Kondrich and Aris Konstantinidis and Kyle Kosic and Gretchen Krueger and Vishal Kuo and Michael Lampe and Ikai Lan and Teddy Lee and Jan Leike and Jade Leung and Daniel Levy and Chak Ming Li and Rachel Lim and Molly Lin and Stephanie Lin and Mateusz Litwin and Theresa Lopez and Ryan Lowe and Patricia Lue and Anna Makanju and Kim Malfacini and Sam Manning and Todor Markov and Yaniv Markovski and Bianca Martin and Katie Mayer and Andrew Mayne and Bob McGrew and Scott Mayer McKinney and Christine McLeavey and Paul McMillan and Jake McNeil and David Medina and Aalok Mehta and Jacob Menick and Luke Metz and Andrey Mishchenko and Pamela Mishkin and Vinnie Monaco and Evan Morikawa and Daniel Mossing and Tong Mu and Mira Murati and Oleg Murk and David Mély and Ashvin Nair and Reiichiro Nakano and Rajeev Nayak and Arvind Neelakantan and Richard Ngo and Hyeonwoo Noh and Long Ouyang and Cullen O'Keefe and Jakub Pachocki and Alex Paino and Joe Palermo and Ashley Pantuliano and Giambattista Parascandolo and Joel Parish and Emy Parparita and Alex Passos and Mikhail Pavlov and Andrew Peng and Adam Perelman and Filipe de Avila Belbute Peres and Michael Petrov and Henrique Ponde de Oliveira Pinto and Michael and Pokorny and Michelle Pokrass and Vitchyr H. Pong and Tolly Powell and Alethea Power and Boris Power and Elizabeth Proehl and Raul Puri and Alec Radford and Jack Rae and Aditya Ramesh and Cameron Raymond and Francis Real and Kendra Rimbach and Carl Ross and Bob Rotsted and Henri Roussez and Nick Ryder and Mario Saltarelli and Ted Sanders and Shibani Santurkar and Girish Sastry and Heather Schmidt and David Schnurr and John Schulman and Daniel Selsam and Kyla Sheppard and Toki Sherbakov and Jessica Shieh and Sarah Shoker and Pranav Shyam and Szymon Sidor and Eric Sigler and Maddie Simens and Jordan Sitkin and Katarina Slama and Ian Sohl and Benjamin Sokolowsky and Yang Song and Natalie Staudacher and Felipe Petroski Such and Natalie Summers and Ilya Sutskever and Jie Tang and Nikolas Tezak and Madeleine B. Thompson and Phil Tillet and Amin Tootoonchian and Elizabeth Tseng and Preston Tuggle and Nick Turley and Jerry Tworek and Juan Felipe Cerón Uribe and Andrea Vallone and Arun Vijayvergiya and Chelsea Voss and Carroll Wainwright and Justin Jay Wang and Alvin Wang and Ben Wang and Jonathan Ward and Jason Wei and CJ Weinmann and Akila Welihinda and Peter Welinder and Jiayi Weng and Lilian Weng and Matt Wiethoff and Dave Willner and Clemens Winter and Samuel Wolrich and Hannah Wong and Lauren Workman and Sherwin Wu and Jeff Wu and Michael Wu and Kai Xiao and Tao Xu and Sarah Yoo and Kevin Yu and Qiming Yuan and Wojciech Zaremba and Rowan Zellers and Chong Zhang and Marvin Zhang and Shengjia Zhao and Tianhao Zheng and Juntang Zhuang and William Zhuk and Barret Zoph},
      year={2024},
      eprint={2303.08774},
      archivePrefix={arXiv},
      primaryClass={cs.CL},
      url={https://arxiv.org/abs/2303.08774}, 
}

@article{holmes2020researcher,
  title={Researcher Positionality--A Consideration of Its Influence and Place in Qualitative Research--A New Researcher Guide.},
  author={Holmes, Andrew Gary Darwin},
  journal={Shanlax International Journal of Education},
  volume={8},
  number={4},
  pages={1--10},
  year={2020},
  publisher={ERIC}
}

@article{iliadis2016critical,
  title={Critical data studies: An introduction},
  author={Iliadis, Andrew and Russo, Federica},
  journal={Big Data \& Society},
  volume={3},
  number={2},
  pages={2053951716674238},
  year={2016},
  publisher={SAGE Publications Sage UK: London, England}
}

@article{braun2006using,
  title={Using thematic analysis in psychology},
  author={Braun, Virginia and Clarke, Victoria},
  journal={Qualitative research in psychology},
  volume={3},
  number={2},
  pages={77--101},
  year={2006},
  publisher={Taylor \& Francis}
}

@article{souly2024strongreject,
  title={A strongreject for empty jailbreaks},
  author={Souly, Alexandra and Lu, Qingyuan and Bowen, Dillon and Trinh, Tu and Hsieh, Elvis and Pandey, Sana and Abbeel, Pieter and Svegliato, Justin and Emmons, Scott and Watkins, Olivia and others},
  journal={Advances in Neural Information Processing Systems},
  volume={37},
  pages={125416--125440},
  year={2024}
}

@inproceedings{birhane2022values,
  title={The values encoded in machine learning research},
  author={Birhane, Abeba and Kalluri, Pratyusha and Card, Dallas and Agnew, William and Dotan, Ravit and Bao, Michelle},
  booktitle={Proceedings of the 2022 ACM conference on fairness, accountability, and transparency},
  pages={173--184},
  year={2022}
}

@article{lin2025against,
  title={Against The Achilles' Heel: A Survey on Red Teaming for Generative Models},
  author={Lin, Lizhi and Mu, Honglin and Zhai, Zenan and Wang, Minghan and Wang, Yuxia and Wang, Renxi and Gao, Junjie and Zhang, Yixuan and Che, Wanxiang and Baldwin, Timothy and others},
  journal={Journal of Artificial Intelligence Research},
  volume={82},
  pages={687--775},
  year={2025}
}

@book{zenko2015red,
  title={Red Team: How to succeed by thinking like the enemy},
  author={Zenko, Micah},
  year={2015},
  publisher={Basic Books}
}

@misc{Anderson2025Jul,
	author = {Anderson, Evan and Holdsworth, Jim and Kosinski, Matthew},
	title = {{What is Red teaming}},
	year = {2025},
	month = jul,
	note = {[Online; accessed 27. Jul. 2025]},
	url = {https://www.ibm.com/think/topics/red-teaming}
}

@article{Solaiman2023Jun,
	author = {Solaiman, Irene and Talat, Zeerak and Agnew, William and Ahmad, Lama and Baker, Dylan and Blodgett, Su Lin and Chen, Canyu and Daum{\ifmmode\acute{e}\else\'{e}\fi} Iii, Hal and Dodge, Jesse and Duan, Isabella and Evans, Ellie and Friedrich, Felix and Ghosh, Avijit and Gohar, Usman and Hooker, Sara and Jernite, Yacine and Kalluri, Ria and Lusoli, Alberto and Leidinger, Alina and Lin, Michelle and Lin, Xiuzhu and Luccioni, Sasha and Mickel, Jennifer and Mitchell, Margaret and Newman, Jessica and Ovalle, Anaelia and Png, Marie-Therese and Singh, Shubham and Strait, Andrew and Struppek, Lukas and Subramonian, Arjun},
	title = {{Evaluating the Social Impact of Generative AI Systems in Systems and Society}},
	journal = {arXiv},
	year = {2023},
	month = jun,
	eprint = {2306.05949},
	doi = {10.48550/arXiv.2306.05949}
}

@article{singh2025red,
  title={Red-Teaming in the Public Interest},
  author={Singh, Ranjit and Blili-Hamelin, Borhane and Anderson, Carol and Tafesse, Emnet and Vecchione, Briana and Duckles, Beth and Metcalf, Jacob},
  journal={New York: Data \& Society Research Institute},
  year={2025}
}

@article{Longpre2024Mar,
	author = {Longpre, Shayne and Kapoor, Sayash and Klyman, Kevin and Ramaswami, Ashwin and Bommasani, Rishi and Blili-Hamelin, Borhane and Huang, Yangsibo and Skowron, Aviya and Yong, Zheng-Xin and Kotha, Suhas and Zeng, Yi and Shi, Weiyan and Yang, Xianjun and Southen, Reid and Robey, Alexander and Chao, Patrick and Yang, Diyi and Jia, Ruoxi and Kang, Daniel and Pentland, Sandy and Narayanan, Arvind and Liang, Percy and Henderson, Peter},
	title = {{A Safe Harbor for AI Evaluation and Red Teaming}},
	journal = {arXiv},
	year = {2024},
	month = mar,
	eprint = {2403.04893},
	doi = {10.48550/arXiv.2403.04893}
}

@article{mansfield2018best,
  title={The best form of defence--the benefits of red teaming},
  author={Mansfield-Devine, Steve},
  journal={Computer Fraud \& Security},
  volume={2018},
  number={10},
  pages={8--12},
  year={2018},
  publisher={MA Business London}
}

@misc{openAI2024,
	title = {{GPT-4o System Card}},
	year = {2024},
        author = {{OpenAI}},
	month = jul,
        url = {https://openai.com/index/gpt-4o-system-card}
}

@misc{anthropic2023ModelCard,
  title        = {{Model Card and Evaluations for Claude Models}},
  author       = {{Anthropic}},
  year         = {2023},
  howpublished = {Technical model card},
  note         = {\url{https://www-cdn.anthropic.com/bd2a28d2535bfb0494cc8e2a3bf135d2e7523226/Model-Card-Claude-2.pdf}},
}

@article{Majumdar2025Jul,
	author = {Majumdar, Subhabrata and Pendleton, Brian and Gupta, Abhishek},
	title = {{Red Teaming AI Red Teaming}},
	journal = {arXiv},
	year = {2025},
	month = jul,
	eprint = {2507.05538},
	doi = {10.48550/arXiv.2507.05538}
}

@article{Liang2022Nov,
	author = {Liang, Percy and Bommasani, Rishi and Lee, Tony and Tsipras, Dimitris and Soylu, Dilara and Yasunaga, Michihiro and Zhang, Yian and Narayanan, Deepak and Wu, Yuhuai and Kumar, Ananya and Newman, Benjamin and Yuan, Binhang and Yan, Bobby and Zhang, Ce and Cosgrove, Christian and Manning, Christopher D. and R{\ifmmode\acute{e}\else\'{e}\fi}, Christopher and Acosta-Navas, Diana and Hudson, Drew A. and Zelikman, Eric and Durmus, Esin and Ladhak, Faisal and Rong, Frieda and Ren, Hongyu and Yao, Huaxiu and Wang, Jue and Santhanam, Keshav and Orr, Laurel and Zheng, Lucia and Yuksekgonul, Mert and Suzgun, Mirac and Kim, Nathan and Guha, Neel and Chatterji, Niladri and Khattab, Omar and Henderson, Peter and Huang, Qian and Chi, Ryan and Xie, Sang Michael and Santurkar, Shibani and Ganguli, Surya and Hashimoto, Tatsunori and Icard, Thomas and Zhang, Tianyi and Chaudhary, Vishrav and Wang, William and Li, Xuechen and Mai, Yifan and Zhang, Yuhui and Koreeda, Yuta},
	title = {{Holistic Evaluation of Language Models}},
	journal = {arXiv},
	year = {2022},
	month = nov,
	eprint = {2211.09110},
	doi = {10.48550/arXiv.2211.09110}
}

@article{Perez2022Feb,
	author = {Perez, Ethan and Huang, Saffron and Song, Francis and Cai, Trevor and Ring, Roman and Aslanides, John and Glaese, Amelia and McAleese, Nat and Irving, Geoffrey},
	title = {{Red Teaming Language Models with Language Models}},
	journal = {arXiv},
	year = {2022},
	month = feb,
	eprint = {2202.03286},
	doi = {10.48550/arXiv.2202.03286}
}

@article{Ganguli2022Aug,
	author = {Ganguli, Deep and Lovitt, Liane and Kernion, Jackson and Askell, Amanda and Bai, Yuntao and Kadavath, Saurav and Mann, Ben and Perez, Ethan and Schiefer, Nicholas and Ndousse, Kamal and Jones, Andy and Bowman, Sam and Chen, Anna and Conerly, Tom and DasSarma, Nova and Drain, Dawn and Elhage, Nelson and El-Showk, Sheer and Fort, Stanislav and Hatfield-Dodds, Zac and Henighan, Tom and Hernandez, Danny and Hume, Tristan and Jacobson, Josh and Johnston, Scott and Kravec, Shauna and Olsson, Catherine and Ringer, Sam and Tran-Johnson, Eli and Amodei, Dario and Brown, Tom and Joseph, Nicholas and McCandlish, Sam and Olah, Chris and Kaplan, Jared and Clark, Jack},
	title = {{Red Teaming Language Models to Reduce Harms: Methods, Scaling Behaviors, and Lessons Learned}},
	journal = {arXiv},
	year = {2022},
	month = aug,
	eprint = {2209.07858},
	doi = {10.48550/arXiv.2209.07858}
}

@incollection{Feffer2024Oct,
	author = {Feffer, Michael and Sinha, Anusha and Deng, Wesley H. and Lipton, Zachary C. and Heidari, Hoda},
	title = {{Red-Teaming for Generative AI: Silver Bullet or Security Theater?}},
	booktitle = {{ACM Conferences}},
	pages = {421--437},
	year = {2024},
	month = oct,
	publisher = {AAAI Press},
	doi = {https://doi.org/10.1609/aies.v7i1.31647}
}

@article{Weidinger2023eval,
	author = {Weidinger, Laura and Rauh, Maribeth and Marchal, Nahema and Manzini, Arianna and Hendricks, Lisa Anne and Mateos-Garcia, Juan and Bergman, Stevie and Kay, Jackie and Griffin, Conor and Bariach, Ben and Gabriel, Iason and Rieser, Verena and Isaac, William},
	title = {{Sociotechnical Safety Evaluation of Generative AI Systems}},
	journal = {arXiv},
	year = {2023},
	month = oct,
	eprint = {2310.11986},
	doi = {10.48550/arXiv.2310.11986}
}

@article{Sinha2025Jul,
	author = {Sinha, Anusha and Lucassen, James and Grimes, Keltin and Feffer, Michael and Soto, Mary and Heidari, Hoda and Vanhoudnos, Nathan},
	title = {{What Can Generative AI Red-Teaming Learn from Cyber Red-Teaming?}},
	year = {2025},
	month = jul,
	publisher = {Carnegie Mellon University},
	doi = {10.1184/R1/29410136.v1}
}

@incollection{Xiao2024audit,
	author = {Xiao, Ziang and Deng, Wesley Hanwen and Lam, Michelle S. and Eslami, Motahhare and Kim, Juho and Lee, Mina and Liao, Q. Vera},
	title = {{Human-Centered Evaluation and Auditing of Language Models}},
	booktitle = {{ACM Conferences}},
	pages = {1--6},
	year = {2024},
	month = may,
	publisher = {Association for Computing Machinery},
	address = {New York, NY, USA},
	doi = {10.1145/3613905.3636302}
}

@article{Zhang2025Jun,
	author = {Zhang, Alice Qian and Suh, Jina and Gray, Mary L. and Shen, Hong},
	title = {{Effective Automation to Support the Human Infrastructure in AI Red Teaming}},
	journal = {interactions},
	volume = {32},
	number = {4},
	pages = {58--61},
	year = {2025},
	month = jun,
	issn = {1072-5520},
	publisher = {Association for Computing Machinery},
	doi = {10.1145/3731866}
}

@article{Gillespie2024Dec,
	author = {Gillespie, Tarleton and Shaw, Ryland and Gray, Mary L. and Suh, Jina},
	title = {{AI red-teaming is a sociotechnical challenge: on values, labor, and harms}},
	journal = {arXiv},
	year = {2024},
	month = dec,
	eprint = {2412.09751},
	doi = {10.48550/arXiv.2412.09751}
}

@book{Costanza-Chock2020Mar,
	author = {Costanza-Chock, Sasha},
	title = {{Design Justice: Community-Led Practices to Build the Worlds We Need}},
	journal = {MIT Press},
	year = {2020},
	month = mar,
	isbn = {978-0-26235686-2},
	publisher = {The MIT Press},
	address = {Cambridge, MA, USA},
	doi = {10.7551/mitpress/12255.001.0001}
}

@article{solyst2023bias,
author = {Solyst, Jaemarie and Yang, Ellia and Xie, Shixian and Ogan, Amy and Hammer, Jessica and Eslami, Motahhare},
title = {The Potential of Diverse Youth as Stakeholders in Identifying and Mitigating Algorithmic Bias for a Future of Fairer AI},
year = {2023},
issue_date = {October 2023},
publisher = {Association for Computing Machinery},
address = {New York, NY, USA},
volume = {7},
number = {CSCW2},
url = {https://doi-org.proxy.library.nd.edu/10.1145/3610213},
doi = {10.1145/3610213},
journal = {Proc. ACM Hum.-Comput. Interact.},
month = oct,
articleno = {364},
numpages = {27}
}

@inproceedings{delgado2023participate,
author = {Delgado, Fernando and Yang, Stephen and Madaio, Michael and Yang, Qian},
title = {The Participatory Turn in AI Design: Theoretical Foundations and the Current State of Practice},
year = {2023},
isbn = {9798400703812},
publisher = {Association for Computing Machinery},
address = {New York, NY, USA},
url = {https://doi-org.proxy.library.nd.edu/10.1145/3617694.3623261},
doi = {10.1145/3617694.3623261},
booktitle = {Proceedings of the 3rd ACM Conference on Equity and Access in Algorithms, Mechanisms, and Optimization},
articleno = {37},
numpages = {23},
location = {Boston, MA, USA},
series = {EAAMO '23}
}

@incollection{Morales2024auditor,
	author = {Morales-Navarro, Luis and Kafai, Yasmin and Konda, Vedya and Metaxa, Dana{\ifmmode\ddot{e}\else\"{e}\fi}},
	title = {{Youth as Peer Auditors: Engaging Teenagers with Algorithm Auditing of Machine Learning Applications}},
	booktitle = {{ACM Conferences}},
	pages = {560--573},
	year = {2024},
	month = jun,
	publisher = {Association for Computing Machinery},
	address = {New York, NY, USA},
	doi = {10.1145/3628516.3655752}
}

@misc{cais2023statement,
  author       = {{Center for AI Safety}},
  title        = {Statement on AI Risk},
  year         = {2023},
  howpublished = {\url{https://www.safe.ai/statement-on-ai-risk}},
}

@incollection{Qian2024May,
	author = {Qian, Crystal and Reif, Emily and Kahng, Minsuk},
	title = {{Understanding the Dataset Practitioners Behind Large Language Models}},
	booktitle = {{ACM Conferences}},
	pages = {1--7},
	year = {2024},
	month = may,
	publisher = {Association for Computing Machinery},
	address = {New York, NY, USA},
	doi = {10.1145/3613905.3651007}
}

@misc{EUAIact,
	author = {{European Parliament}},
	title = {{EU AI Act: first regulation on artificial intelligence {$\vert$} Topics {$\vert$} European Parliament}},
	journal = {Topics {$\vert$} European Parliament},
	year = {2025},
	month = aug,
	url = {https://www.europarl.europa.eu/topics/en/article/20230601STO93804/eu-ai-act-first-regulation-on-artificial-intelligence}
}

@article{Oguine2025Apr,
author = {Oguine, Ozioma Collins and Anuyah, Oghenemaro and Agha, Zainab and Melgarez, Iris and Alvarado Garcia, Adriana and Badillo-Urquiola, Karla},
title = {Online Safety for All: Sociocultural Insights from a Systematic Review of Youth Online Safety in the Global South},
year = {2025},
issue_date = {November 2025},
publisher = {Association for Computing Machinery},
address = {New York, NY, USA},
volume = {9},
number = {7},
url = {https://doi.org/10.1145/3757639},
doi = {10.1145/3757639},
journal = {Proc. ACM Hum.-Comput. Interact.},
month = oct,
articleno = {CSCW458},
numpages = {30}
}

@article{Badillo-Urquiola2024Apr,
	author = {Badillo-Urquiola, Karla and Agha, Zainab and Abaquita, Denielle and Harpin, Scott B. and Wisniewski, Pamela J.},
	title = {{Towards a Social Ecological Approach to Supporting Caseworkers in Promoting the Online Safety of Youth in Foster Care}},
	journal = {Proc. ACM Hum.-Comput. Interact.},
	volume = {8},
	number = {CSCW1},
	pages = {1--28},
	year = {2024},
	month = apr,
	publisher = {Association for Computing Machinery},
	doi = {10.1145/3637412}
}

@article{Weidinger2021harms,
	author = {Weidinger, Laura and Mellor, John and Rauh, Maribeth and Griffin, Conor and Uesato, Jonathan and Huang, Po-Sen and Cheng, Myra and Glaese, Mia and Balle, Borja and Kasirzadeh, Atoosa and Kenton, Zac and Brown, Sasha and Hawkins, Will and Stepleton, Tom and Biles, Courtney and Birhane, Abeba and Haas, Julia and Rimell, Laura and Hendricks, Lisa Anne and Isaac, William and Legassick, Sean and Irving, Geoffrey and Gabriel, Iason},
	title = {{Ethical and social risks of harm from Language Models}},
	journal = {arXiv},
	year = {2021},
	month = dec,
	eprint = {2112.04359},
	doi = {10.48550/arXiv.2112.04359}
}

@incollection{Muller2022forgetting,
	author = {Muller, Michael and Strohmayer, Angelika},
	title = {{Forgetting Practices in the Data Sciences}},
	booktitle = {{ACM Conferences}},
	pages = {1--19},
	year = {2022},
	month = apr,
	publisher = {Association for Computing Machinery},
	address = {New York, NY, USA},
	doi = {10.1145/3491102.3517644}
}

@incollection{He2024AI-collaborate,
	author = {He, Jessica and Houde, Stephanie and Gonzalez, Gabriel E. and Silva Moran, Dar{\ifmmode\acute{\imath}\else\'{\i}\fi}o Andr{\ifmmode\acute{e}\else\'{e}\fi}s and Ross, Steven I. and Muller, Michael and Weisz, Justin D.},
	title = {{AI and the Future of Collaborative Work: Group Ideation with an LLM in a Virtual Canvas}},
	booktitle = {{ACM Other conferences}},
	pages = {1--14},
	year = {2024},
	month = jun,
	publisher = {Association for Computing Machinery},
	address = {New York, NY, USA},
	doi = {10.1145/3663384.3663398}
}

@article{Wong2017Dec,
	author = {Wong, Richmond Y. and Mulligan, Deirdre K. and Van Wyk, Ellen and Pierce, James and Chuang, John},
	title = {{Eliciting Values Reflections by Engaging Privacy Futures Using Design Workbooks}},
	journal = {Proc. ACM Hum.-Comput. Interact.},
	volume = {1},
	number = {CSCW},
	pages = {1--26},
	year = {2017},
	month = dec,
	publisher = {Association for Computing Machinery},
	doi = {10.1145/3134746}
}

@article{Biden2023AI,
	author = {The White House},
	title = {{FACT SHEET: President Biden Issues Executive Order on Safe, Secure, and Trustworthy Artificial Intelligence {$\vert$} The White House}},
	journal = {White House},
	year = {2023},
	month = oct,
	url = {https://bidenwhitehouse.archives.gov/briefing-room/statements-releases/2023/10/30/fact-sheet-president-biden-issues-executive-order-on-safe-secure-and-trustworthy-artificial-intelligence/}
}

@incollection{Sambasivan2021May,
	author = {Sambasivan, Nithya and Kapania, Shivani and Highfill, Hannah and Akrong, Diana and Paritosh, Praveen and Aroyo, Lora M.},
	title = {{{\textquotedblleft}Everyone wants to do the model work, not the data work{\textquotedblright}: Data Cascades in High-Stakes AI}},
	booktitle = {{ACM Conferences}},
	pages = {1--15},
	year = {2021},
	month = may,
	publisher = {Association for Computing Machinery},
	address = {New York, NY, USA},
	doi = {10.1145/3411764.3445518}
}

@incollection{AlvaradoGarcia2025Apr,
	author = {Alvarado Garcia, Adriana and Candello, Heloisa and Badillo-Urquiola, Karla and Wong-Villacres, Marisol},
	title = {{Emerging Data Practices: Data Work in the Era of Large Language Models}},
	booktitle = {{ACM Conferences}},
	pages = {1--21},
	year = {2025},
	month = apr,
	publisher = {Association for Computing Machinery},
	address = {New York, NY, USA},
	doi = {10.1145/3706598.3714069}
}

@incollection{Selbst2019Jan,
	author = {Selbst, Andrew D. and Boyd, Danah and Friedler, Sorelle A. and Venkatasubramanian, Suresh and Vertesi, Janet},
	title = {{Fairness and Abstraction in Sociotechnical Systems}},
	booktitle = {{ACM Conferences}},
	pages = {59--68},
	year = {2019},
	month = jan,
	publisher = {Association for Computing Machinery},
	address = {New York, NY, USA},
	doi = {10.1145/3287560.3287598}
}

@incollection{Applebaum2016redteam,
	author = {Applebaum, Andy and Miller, Doug and Strom, Blake and Korban, Chris and Wolf, Ross},
	title = {{Intelligent, automated red team emulation}},
	booktitle = {{ACM Other conferences}},
	pages = {363--373},
	year = {2016},
	month = dec,
	publisher = {Association for Computing Machinery},
	address = {New York, NY, USA},
	doi = {10.1145/2991079.2991111}
}

@article{Ahmad2025openai,
	author = {Ahmad, Lama and Agarwal, Sandhini and Lampe, Michael and Mishkin, Pamela},
	title = {{OpenAI's Approach to External Red Teaming for AI Models and Systems}},
	journal = {arXiv},
	year = {2025},
	month = jan,
	eprint = {2503.16431},
	doi = {10.48550/arXiv.2503.16431}
}

@article{Gehman2020Toxicity,
	author = {Gehman, Samuel and Gururangan, Suchin and Sap, Maarten and Choi, Yejin and Smith, Noah A.},
	title = {{RealToxicityPrompts: Evaluating Neural Toxic Degeneration in Language Models}},
	journal = {ACL Anthology},
	pages = {3356--3369},
	year = {2020},
	month = nov,
	doi = {10.18653/v1/2020.findings-emnlp.301}
}

@article{Sheng2019Nov,
	author = {Sheng, Emily and Chang, Kai-Wei and Natarajan, Prem and Peng, Nanyun},
	title = {{The Woman Worked as a Babysitter: On Biases in Language Generation}},
	journal = {ACL Anthology},
	pages = {3407--3412},
	year = {2019},
	month = nov,
	doi = {10.18653/v1/D19-1339}
}

@article{Welbl2021Nov,
	author = {Welbl, Johannes and Glaese, Amelia and Uesato, Jonathan and Dathathri, Sumanth and Mellor, John and Hendricks, Lisa Anne and Anderson, Kirsty and Kohli, Pushmeet and Coppin, Ben and Huang, Po-Sen},
	title = {{Challenges in Detoxifying Language Models}},
	journal = {ACL Anthology},
	pages = {2447--2469},
	year = {2021},
	month = nov,
	doi = {10.18653/v1/2021.findings-emnlp.210}
}

@article{Rajpurkar2016Nov,
	author = {Rajpurkar, Pranav and Zhang, Jian and Lopyrev, Konstantin and Liang, Percy},
	title = {{SQuAD: 100,000+ Questions for Machine Comprehension of Text}},
	journal = {ACL Anthology},
	pages = {2383--2392},
	year = {2016},
	month = nov,
	doi = {10.18653/v1/D16-1264}
}

@article{Hofmann2024Sep,
	author = {Hofmann, Valentin and Kalluri, Pratyusha Ria and Jurafsky, Dan and King, Sharese},
	title = {{AI generates covertly racist decisions about people based on their dialect}},
	journal = {Nature},
	volume = {633},
	pages = {147--154},
	year = {2024},
	month = sep,
	issn = {1476-4687},
	publisher = {Nature Publishing Group},
	doi = {10.1038/s41586-024-07856-5}
}

@inproceedings{oguine2025bridging,
  author       = {Oguine, Ozioma C. and Olesk, Johanna and Solyst, Jaemarie and Madaio, Michael and Muller, Michael and Alvarado Garcia, Adriana and Badillo-Urquiola, Karla},
  title        = {Bridging Expertise and Participation in AI: Multistakeholder Approaches to Safer AI Systems for Youth Online Safety},
  booktitle    = {Companion of the 2025 ACM Conference on Computer-Supported Cooperative Work and Social Computing (CSCW Companion ’25)},
  year         = {2025},
  pages        = {6 pages},
  address      = {Bergen, Norway},
  publisher    = {ACM},
  doi          = {10.1145/3715070.3748294},
  url          = {https://doi.org/10.1145/3715070.3748294}
}

@article{Wang2018Nov,
	author = {Wang, Alex and Singh, Amanpreet and Michael, Julian and Hill, Felix and Levy, Omer and Bowman, Samuel},
	title = {{GLUE: A Multi-Task Benchmark and Analysis Platform for Natural Language Understanding}},
	journal = {ACL Anthology},
	pages = {353--355},
	year = {2018},
	month = nov,
	doi = {10.18653/v1/W18-5446}
}

@incollection{Kallina2025Jun,
	author = {Kallina, Emma and Bohn{\ifmmode\acute{e}\else\'{e}\fi}, Thomas and Singh, Jatinder},
	title = {{Stakeholder Participation for Responsible AI Development: Disconnects Between Guidance and Current Practice}},
	booktitle = {{ACM Conferences}},
	pages = {1060--1079},
	year = {2025},
	month = jun,
	publisher = {Association for Computing Machinery},
	address = {New York, NY, USA},
	doi = {10.1145/3715275.3732069}
}

@inproceedings{Weidinger2022,
author = {Weidinger, Laura and Uesato, Jonathan and Rauh, Maribeth and Griffin, Conor and Huang, Po-Sen and Mellor, John and Glaese, Amelia and Cheng, Myra and Balle, Borja and Kasirzadeh, Atoosa and Biles, Courtney and Brown, Sasha and Kenton, Zac and Hawkins, Will and Stepleton, Tom and Birhane, Abeba and Hendricks, Lisa Anne and Rimell, Laura and Isaac, William and Haas, Julia and Legassick, Sean and Irving, Geoffrey and Gabriel, Iason},
title = {Taxonomy of Risks posed by Language Models},
year = {2022},
isbn = {9781450393522},
publisher = {Association for Computing Machinery},
address = {New York, NY, USA},
url = {https://doi.org/10.1145/3531146.3533088},
doi = {10.1145/3531146.3533088},
booktitle = {Proceedings of the 2022 ACM Conference on Fairness, Accountability, and Transparency},
pages = {214–229},
numpages = {16},
location = {Seoul, Republic of Korea},
series = {FAccT '22}
}

@article{Zou2023universal,
	author = {Zou, Andy and Wang, Zifan and Carlini, Nicholas and Nasr, Milad and Kolter, J. Zico and Fredrikson, Matt},
	title = {{Universal and Transferable Adversarial Attacks on Aligned Language Models}},
	journal = {arXiv},
	year = {2023},
	month = jul,
	eprint = {2307.15043},
	doi = {10.48550/arXiv.2307.15043}
}

@article{Mazeika2024harmbench,
	author = {Mazeika, Mantas and Phan, Long and Yin, Xuwang and Zou, Andy and Wang, Zifan and Mu, Norman and Sakhaee, Elham and Li, Nathaniel and Basart, Steven and Li, Bo and Forsyth, David and Hendrycks, Dan},
	title = {{HarmBench: A Standardized Evaluation Framework for Automated Red Teaming and Robust Refusal}},
	journal = {arXiv},
	year = {2024},
	month = feb,
	eprint = {2402.04249},
	doi = {10.48550/arXiv.2402.04249}
}

@article{Ji2023Beavertails,
	author = {Ji, Jiaming and Liu, Mickel and Dai, Juntao and Pan, Xuehai and Zhang, Chi and Bian, Ce and Zhang, Chi and Sun, Ruiyang and Wang, Yizhou and Yang, Yaodong},
	title = {{BeaverTails: Towards Improved Safety Alignment of LLM via a Human-Preference Dataset}},
	journal = {arXiv},
	year = {2023},
	month = jul,
	eprint = {2307.04657},
	doi = {10.48550/arXiv.2307.04657}
}

@article{Miceli2020,
author = {Miceli, Milagros and Schuessler, Martin and Yang, Tianling},
title = {Between Subjectivity and Imposition: Power Dynamics in Data Annotation for Computer Vision},
year = {2020},
issue_date = {October 2020},
publisher = {Association for Computing Machinery},
address = {New York, NY, USA},
volume = {4},
number = {CSCW2},
url = {https://doi.org/10.1145/3415186},
doi = {10.1145/3415186},
journal = {Proc. ACM Hum.-Comput. Interact.},
month = oct,
articleno = {115},
numpages = {25}
}

@article{Scheuerman2021,
author = {Scheuerman, Morgan Klaus and Hanna, Alex and Denton, Remi},
title = {Do Datasets Have Politics? Disciplinary Values in Computer Vision Dataset Development},
year = {2021},
issue_date = {October 2021},
publisher = {Association for Computing Machinery},
address = {New York, NY, USA},
volume = {5},
number = {CSCW2},
url = {https://doi.org/10.1145/3476058},
doi = {10.1145/3476058},
journal = {Proc. ACM Hum.-Comput. Interact.},
month = oct,
articleno = {317},
numpages = {37}
}

@inproceedings{chen_2023,
author = {Chen, Bocheng and Wang, Guangjing and Guo, Hanqing and Wang, Yuanda and Yan, Qiben},
title = {Understanding Multi-Turn Toxic Behaviors in Open-Domain Chatbots},
year = {2023},
isbn = {9798400707650},
publisher = {Association for Computing Machinery},
address = {New York, NY, USA},
url = {https://doi.org/10.1145/3607199.3607237},
doi = {10.1145/3607199.3607237},
booktitle = {Proceedings of the 26th International Symposium on Research in Attacks, Intrusions and Defenses},
pages = {282–296},
numpages = {15},
location = {Hong Kong, China},
series = {RAID '23}
}

@misc{Martone_2025, 
title={Ai is sexually harassing our kids. here’s how legislators can stop it.}, 
url={https://www.techpolicy.press/ai-is-sexually-harassing-our-kids-heres-how-legislators-can-stop-it/}, journal={Tech Policy Press}, 
publisher={Tech Policy Press}, 
author={Martone, Omny Miranda}, 
year={2025}, month={Aug}}

@misc{yu2025,
      title={Understanding Generative AI Risks for Youth: A Taxonomy Based on Empirical Data}, 
      author={Yaman Yu and Yiren Liu and Jacky Zhang and Yun Huang and Yang Wang},
      year={2025},
      eprint={2502.16383},
      archivePrefix={arXiv},
      primaryClass={cs.HC},
      url={https://arxiv.org/abs/2502.16383}, 
}

@misc{schoene2025,
      title={`For Argument's Sake, Show Me How to Harm Myself!': Jailbreaking LLMs in Suicide and Self-Harm Contexts}, 
      author={Annika M Schoene and Cansu Canca},
      year={2025},
      eprint={2507.02990},
      archivePrefix={arXiv},
      primaryClass={cs.CL},
      url={https://arxiv.org/abs/2507.02990}, 
}

@inproceedings{xu-etal-2021-bot,
    title = "Bot-Adversarial Dialogue for Safe Conversational Agents",
    author = "Xu, Jing  and
      Ju, Da  and
      Li, Margaret  and
      Boureau, Y-Lan  and
      Weston, Jason  and
      Dinan, Emily",
    editor = "Toutanova, Kristina  and
      Rumshisky, Anna  and
      Zettlemoyer, Luke  and
      Hakkani-Tur, Dilek  and
      Beltagy, Iz  and
      Bethard, Steven  and
      Cotterell, Ryan  and
      Chakraborty, Tanmoy  and
      Zhou, Yichao",
    booktitle = "Proceedings of the 2021 Conference of the North American Chapter of the Association for Computational Linguistics: Human Language Technologies",
    month = jun,
    year = "2021",
    address = "Online",
    publisher = "Association for Computational Linguistics",
    url = "https://aclanthology.org/2021.naacl-main.235/",
    doi = "10.18653/v1/2021.naacl-main.235",
    pages = "2950--2968"
}

@misc{schwartz2025,
      title={Reality Check: A New Evaluation Ecosystem Is Necessary to Understand AI's Real World Effects}, 
      author={Reva Schwartz and Rumman Chowdhury and Akash Kundu and Heather Frase and Marzieh Fadaee and Tom David and Gabriella Waters and Afaf Taik and Morgan Briggs and Patrick Hall and Shomik Jain and Kyra Yee and Spencer Thomas and Sundeep Bhandari and Paul Duncan and Andrew Thompson and Maya Carlyle and Qinghua Lu and Matthew Holmes and Theodora Skeadas},
      year={2025},
      eprint={2505.18893},
      archivePrefix={arXiv},
      primaryClass={cs.CY},
      url={https://arxiv.org/abs/2505.18893}, 
}

@misc{reiter2025,
      title={We Should Evaluate Real-World Impact}, 
      author={Ehud Reiter},
      year={2025},
      eprint={2507.05973},
      archivePrefix={arXiv},
      primaryClass={cs.CL},
      url={https://arxiv.org/abs/2507.05973}, 
}

@misc{burden2024,
      title={Evaluating AI Evaluation: Perils and Prospects}, 
      author={John Burden},
      year={2024},
      eprint={2407.09221},
      archivePrefix={arXiv},
      primaryClass={cs.AI},
      url={https://arxiv.org/abs/2407.09221}, 
}

@misc{francois2025,
      title={A Different Approach to AI Safety: Proceedings from the Columbia Convening on Openness in Artificial Intelligence and AI Safety}, 
      author={Camille François and Ludovic Péran and Ayah Bdeir and Nouha Dziri and Will Hawkins and Yacine Jernite and Sayash Kapoor and Juliet Shen and Heidy Khlaaf and Kevin Klyman and Nik Marda and Marie Pellat and Deb Raji and Divya Siddarth and Aviya Skowron and Joseph Spisak and Madhulika Srikumar and Victor Storchan and Audrey Tang and Jen Weedon},
      year={2025},
      eprint={2506.22183},
      archivePrefix={arXiv},
      primaryClass={cs.AI},
      url={https://arxiv.org/abs/2506.22183}, 
}

@inproceedings{liao2021,
 author = {Liao, Thomas and Taori, Rohan and Raji, Deborah and Schmidt, Ludwig},
 booktitle = {Proceedings of the Neural Information Processing Systems Track on Datasets and Benchmarks},
 editor = {J. Vanschoren and S. Yeung},
 pages = {},
 title = {Are We Learning Yet? A Meta Review of Evaluation Failures Across Machine Learning},
 url = {https://datasets-benchmarks-proceedings.neurips.cc/paper_files/paper/2021/file/757b505cfd34c64c85ca5b5690ee5293-Paper-round2.pdf},
 volume = {1},
 year = {2021}
}

@misc{eriksson2025,
      title={Can We Trust AI Benchmarks? An Interdisciplinary Review of Current Issues in AI Evaluation}, 
      author={Maria Eriksson and Erasmo Purificato and Arman Noroozian and Joao Vinagre and Guillaume Chaslot and Emilia Gomez and David Fernandez-Llorca},
      year={2025},
      eprint={2502.06559},
      archivePrefix={arXiv},
      primaryClass={cs.AI},
      url={https://arxiv.org/abs/2502.06559}, 
}

@article{Dourish_2021,
author = {Dourish, Paul},
title = {Seeking a foundation for context-aware computing},
year = {2001},
issue_date = {December 2001},
publisher = {L. Erlbaum Associates Inc.},
address = {USA},
volume = {16},
number = {2},
issn = {0737-0024},
url = {https://doi.org/10.1207/S15327051HCI16234_07},
doi = {10.1207/S15327051HCI16234_07},
journal = {Hum.-Comput. Interact.},
month = dec,
pages = {229–241},
numpages = {13}
}

@book{suchman_1987,
author = {Suchman, Lucy A.},
title = {Plans and situated actions: the problem of human-machine communication},
year = {1987},
isbn = {0521331374},
publisher = {Cambridge University Press},
address = {USA}
}

@misc{casper2023,
      title={Explore, Establish, Exploit: Red Teaming Language Models from Scratch}, 
      author={Stephen Casper and Jason Lin and Joe Kwon and Gatlen Culp and Dylan Hadfield-Menell},
      year={2023},
      eprint={2306.09442},
      archivePrefix={arXiv},
      primaryClass={cs.CL},
      url={https://arxiv.org/abs/2306.09442}, 
}

@article{Dey_2001,
author = {Dey, Anind K.},
title = {Understanding and Using Context},
year = {2001},
issue_date = {February 2001},
publisher = {Springer-Verlag},
address = {Berlin, Heidelberg},
volume = {5},
number = {1},
issn = {1617-4909},
url = {https://doi.org/10.1007/s007790170019},
doi = {10.1007/s007790170019},
journal = {Personal Ubiquitous Comput.},
month = jan,
pages = {4–7},
numpages = {4}
}

@InProceedings{abowd_1999,
author="Abowd, Gregory D.
and Dey, Anind K.
and Brown, Peter J.
and Davies, Nigel
and Smith, Mark
and Steggles, Pete",
editor="Gellersen, Hans-W.",
title="Towards a Better Understanding of Context and Context-Awareness",
booktitle="Handheld and Ubiquitous Computing",
year="1999",
publisher="Springer Berlin Heidelberg",
address="Berlin, Heidelberg",
pages="304--307",
isbn="978-3-540-48157-7"
}

@inproceedings{singhania-etal-2025-multi,
    title = "Multi-lingual Multi-turn Automated Red Teaming for {LLM}s",
    author = "Singhania, Abhishek  and
      Dupuy, Christophe  and
      Mangale, Shivam Sadashiv  and
      Namboori, Amani",
    editor = "Cao, Trista  and
      Das, Anubrata  and
      Kumarage, Tharindu  and
      Wan, Yixin  and
      Krishna, Satyapriya  and
      Mehrabi, Ninareh  and
      Dhamala, Jwala  and
      Ramakrishna, Anil  and
      Galystan, Aram  and
      Kumar, Anoop  and
      Gupta, Rahul  and
      Chang, Kai-Wei",
    booktitle = "Proceedings of the 5th Workshop on Trustworthy NLP (TrustNLP 2025)",
    month = may,
    year = "2025",
    address = "Albuquerque, New Mexico",
    publisher = "Association for Computational Linguistics",
    url = "https://aclanthology.org/2025.trustnlp-main.11/",
    doi = "10.18653/v1/2025.trustnlp-main.11",
    pages = "141--154",
    ISBN = "979-8-89176-233-6"
}

@inproceedings{ramesh-etal-2023-fairness,
    title = "Fairness in Language Models Beyond {E}nglish: Gaps and Challenges",
    author = "Ramesh, Krithika  and
      Sitaram, Sunayana  and
      Choudhury, Monojit",
    editor = "Vlachos, Andreas  and
      Augenstein, Isabelle",
    booktitle = "Findings of the Association for Computational Linguistics: EACL 2023",
    month = may,
    year = "2023",
    address = "Dubrovnik, Croatia",
    publisher = "Association for Computational Linguistics",
    url = "https://aclanthology.org/2023.findings-eacl.157/",
    doi = "10.18653/v1/2023.findings-eacl.157",
    pages = "2106--2119"
}

@misc{medhelm,
      title={MedHELM: Holistic Evaluation of Large Language Models for Medical Tasks}, 
      author={Suhana Bedi and Hejie Cui and Miguel Fuentes and Alyssa Unell and Michael Wornow and Juan M. Banda and Nikesh Kotecha and Timothy Keyes and Yifan Mai and Mert Oez and Hao Qiu and Shrey Jain and Leonardo Schettini and Mehr Kashyap and Jason Alan Fries and Akshay Swaminathan and Philip Chung and Fateme Nateghi and Asad Aali and Ashwin Nayak and Shivam Vedak and Sneha S. Jain and Birju Patel and Oluseyi Fayanju and Shreya Shah and Ethan Goh and Dong-han Yao and Brian Soetikno and Eduardo Reis and Sergios Gatidis and Vasu Divi and Robson Capasso and Rachna Saralkar and Chia-Chun Chiang and Jenelle Jindal and Tho Pham and Faraz Ghoddusi and Steven Lin and Albert S. Chiou and Christy Hong and Mohana Roy and Michael F. Gensheimer and Hinesh Patel and Kevin Schulman and Dev Dash and Danton Char and Lance Downing and Francois Grolleau and Kameron Black and Bethel Mieso and Aydin Zahedivash and Wen-wai Yim and Harshita Sharma and Tony Lee and Hannah Kirsch and Jennifer Lee and Nerissa Ambers and Carlene Lugtu and Aditya Sharma and Bilal Mawji and Alex Alekseyev and Vicky Zhou and Vikas Kakkar and Jarrod Helzer and Anurang Revri and Yair Bannett and Roxana Daneshjou and Jonathan Chen and Emily Alsentzer and Keith Morse and Nirmal Ravi and Nima Aghaeepour and Vanessa Kennedy and Akshay Chaudhari and Thomas Wang and Sanmi Koyejo and Matthew P. Lungren and Eric Horvitz and Percy Liang and Mike Pfeffer and Nigam H. Shah},
      year={2025},
      eprint={2505.23802},
      archivePrefix={arXiv},
      primaryClass={cs.CL},
      url={https://arxiv.org/abs/2505.23802}, 
}

@inproceedings{wan2023everyone,
  title={Everyone’s voice matters: Quantifying annotation disagreement using demographic information},
  author={Wan, Ruyuan and Kim, Jaehyung and Kang, Dongyeop},
  booktitle={Proceedings of the AAAI Conference on Artificial Intelligence},
  volume={37},
  number={12},
  pages={14523--14530},
  year={2023}
}

@techreport{CommonSenseMedia2025TalkTrustAndTradeOffs,
  author       = {Robb, Michael B. and Mann,  Supreet},
  title        = {Talk, Trust, and Trade‐Offs: The Role of Children’s Online Safety, Privacy, and Learning in the Digital Age},
  institution  = {Common Sense Media},
  year         = {2025},
  url          = {https://www.commonsensemedia.org/sites/default/files/research/report/talk-trust-and-trade-offs_2025_web.pdf},
  note         = {Research report},
}

@misc{inan2023llamaguard,
      title={Llama Guard: LLM-based Input-Output Safeguard for Human-AI Conversations}, 
      author={Hakan Inan and Kartikeya Upasani and Jianfeng Chi and Rashi Rungta and Krithika Iyer and Yuning Mao and Michael Tontchev and Qing Hu and Brian Fuller and Davide Testuggine and Madian Khabsa},
      year={2023},
      eprint={2312.06674},
      archivePrefix={arXiv},
      primaryClass={cs.CL},
      url={https://arxiv.org/abs/2312.06674}, 
}

@article{ahmed,
author = {Nur Ahmed  and Muntasir Wahed  and Neil C. Thompson },
title = {The growing influence of industry in AI research},
journal = {Science},
volume = {379},
number = {6635},
pages = {884-886},
year = {2023},
doi = {10.1126/science.ade2420},
URL = {https://www.science.org/doi/abs/10.1126/science.ade2420}}

@misc{Ahmed_Thompson_2023, 
title={What should be done about the growing influence of industry in AI research?}, 
url={https://www.brookings.edu/articles/what-should-be-done-about-the-growing-influence-of-industry-in-ai-research}, 
journal={ The Brookings Institution}, 
publisher={ The Brookings Institution}, 
author={Ahmed, Nur and  Thompson, Neil C.}, 
year={2023}, month={Dec}}

@misc{Bajema_2024, 
title={Why are large AI models being red teamed?}, 
url={https://spectrum.ieee.org/red-team-ai-llms}, 
journal={IEEE Spectrum}, publisher={IEEE Spectrum}, 
author={Bajema, Natasha}, year={2024}, month={Mar}}

@book{Davies_Vipra_2025, 
title ={Computing Commons: Designing public compute for people and society},
url={https://www.adalovelaceinstitute.org/report/computing-commons/}, 
journal={Computing Commons}, 
publisher ={Ada Lovelace Institute}, author={Davies, Matt and Vipra, Jai}, year={2025}, month={Feb}}

@article{miceli_2022,
author = {Miceli, Milagros and Posada, Julian and Yang, Tianling},
title = {Studying Up Machine Learning Data: Why Talk About Bias When We Mean Power?},
year = {2022},
issue_date = {January 2022},
publisher = {Association for Computing Machinery},
address = {New York, NY, USA},
volume = {6},
number = {GROUP},
url = {https://doi.org/10.1145/3492853},
doi = {10.1145/3492853},
journal = {Proc. ACM Hum.-Comput. Interact.},
month = jan,
articleno = {34},
numpages = {14}
}

@misc{besiroglu2024,
      title={The Compute Divide in Machine Learning: A Threat to Academic Contribution and Scrutiny?}, 
      author={Tamay Besiroglu and Sage Andrus Bergerson and Amelia Michael and Lennart Heim and Xueyun Luo and Neil Thompson},
      year={2024},
      eprint={2401.02452},
      archivePrefix={arXiv},
      primaryClass={cs.CY},
      url={https://arxiv.org/abs/2401.02452}, 
}

\appendix

\end{document}